\documentclass[aps,prd,twocolumn,showpacs,amsmath,amssymb,nofootinbib,preprintnumbers]{revtex4-1}
\usepackage{graphicx}
\usepackage{float}
\usepackage{color}

\newcommand{\be}{\begin{equation}}
\newcommand{\ee}{\end{equation}}
\newcommand{\ba}{\begin{eqnarray}}
\newcommand{\ea}{\end{eqnarray}}

\newcommand{\beq}{\begin{equation}}
\newcommand{\eeq}{\end{equation}}
\newcommand{\beqa}{\begin{eqnarray}}
\newcommand{\eeqa}{\end{eqnarray}}
\newcommand{\nn}{\nonumber}

\newcommand{\tq}{\mathfrak{q}}

\begin{document}

\title{$P-v$ Criticality in Quasitopological Gravity}

\author{Robie A. Hennigar}
\email{rhenniga@uwaterloo.ca}
\affiliation{Department of Physics and Astronomy, University of Waterloo, Waterloo, Ontario N2L 3G1, Canada}
\author{W. G. Brenna}
\email{wbrenna@uwaterloo.ca}
\affiliation{Department of Physics and Astronomy, University of Waterloo, Waterloo, Ontario N2L 3G1, Canada}
\author{Robert B. Mann}
\email{rbmann@uwaterloo.ca}
\affiliation{Department of Physics and Astronomy, University of Waterloo, Waterloo, Ontario N2L 3G1, Canada}

\begin{abstract}
We investigate the thermodynamic behaviour of AdS quasitopological black hole solutions in the context of
extended thermodynamic phase space, in which the cosmological constant induces a pressure with a conjugate volume.
We find that the third order exact quasitopological solution exhibits features consistent with the
third order Lovelock solutions for  positive quasitopological coupling, including multiple reentrant phase transitions and isolated critical
points.  For negative coupling we find the first instances of both reentrant phase transitions and thermodynamic singularities in five dimensions, along with other modified thermodynamic behaviour  compared to   Einstein-AdS-Gauss Bonnet gravity.
\end{abstract}

\maketitle

\section{Introduction}

 Black hole thermodynamics has maintained its popularity, partly because black holes make for compelling
and elegant thermodynamical systems whose behaviour can be linked to the nature of quantum gravity. Of particular
relevance to this work are the asymptotically Anti de-Sitter (AdS) black holes, which have significance in
various proposed gauge-gravity dualities. 

Extended phase space thermodynamics, in which the cosmological constant $\Lambda$ is regarded as a thermodynamic variable 
\cite{Creighton1995} analogous to pressure \cite{Caldarelli2000, Kastor2009, Dolan2010, Dolan2011a, Dolan2011, Dolan2012, Cvetic2010, Larranaga2011, Larranaga2012, 
Gibbons2012, Kubiznak2012, Gunasekaran2012, Belhaj2012,  Lu2012, Smailagic2012, Hendi2012, Cai2013},   has recently become of considerable interest as the proper venue for a complete thermodynamic description of AdS black holes \cite{Kubiznak2014}. The  mass of an AdS black hole is then understood as the enthalpy of the spacetime \cite{Kastor2009}.  
  It was soon realized that the critical exponents of the 4-dimensional Reissner-Nordstr\"om AdS black hole are the same as those in the van der Waals liquid--gas phase transition  \cite{Kubiznak2012}, completing an analogy explored previously in a more restricted context
\cite{Chamblin1999a,Chamblin1999, Cvetic1999, Cvetic1999a}.  Subsequently a broad range of 
new thermodynamic behaviour was discovered, including reentrant phase transitions, tricritical points, 
and more general van der Waals behaviour with standard critical exponents  \cite{Dolan2013, Dolan2013a, Dolan2013b, Gunasekaran2012, Altamirano2013, Frassino2014,
Zou2013,Zou2014,Ma2013, Ma2014, Wei2014,Mo2014, Mo2014a, Mo2014b,Zhang2014, Liu2014, Liu2014a, Johnson2014a,Rajagopal:2014ewa,Delsate:2014zma}, with an extensive review of these issues in the context of rotating black holes presented in \cite{Altamirano:2014tva}.

One of the most recent discoveries of interest was obtained in 3rd-order Lovelock gravity \cite{Frassino2014,Dolan:2014vba}. 
Multiple-reentrant-phase-transition behaviour as well  an isolated critical point with non-standard critical exponents 
 for a $D=7$ third-order (cubic) Lovelock asymptotically AdS black hole were found.   This
latter phenomenon is the first example of a critical point with non-standard critical exponents obtained in a geometric theory of gravity. It was subsequently shown that it occurs for  topological black holes in any $K=(2k+1)$th-order Lovelock gravity theory, provided the Lovelock couplings are appropriately adjusted \cite{Dolan:2014vba}. 
This corresponds to a  double swallowtail behaviour in the Gibbs free energy, 
giving rise to two first-order transitions between the small and large black hole phases.

Our key interest for this paper is to see if these more recently discovered phenomena can occur in a broader context.
 To this end
we explore the extended phase space thermodynamics of black holes in cubic quasitopological gravity.
Quasitopological gravity was proposed \cite{Myers2010,Myers2010a,Oliva2010} as an extension to the Lovelock \cite{Lovelock1971} higher curvature terms. While the Lovelock terms are topological invariants below a certain critical dimensionality, the
quasitopological terms are still active in some lower dimensions. The trade-off is that when spherical symmetry does not hold, the quasitopological terms produce higher than second order derivative terms in the Einstein field equations.

With cubic quasitopological 
gravity we can obtain field equations in $D=5$ similar to those of cubic Lovelock gravity.
We will examine whether  we can likewise obtain similar thermodynamics but in fewer dimensions. One particularly compelling
notion for a $D=5$ black hole is that the basic idea behind AdS/CFT will link it to a $D=4$
field theory, and so perhaps universality relationships with well understood $D=4$ thermodynamic
systems will become apparent.  We find that indeed new features are present in $D=5$, particularly the   first instance of reentrant phase transitions, a phenomenon so far seen only in $D\geq 6$  \cite{Altamirano2013}.

We consider quasitopological additions corresponding to $3^{rd}$-order curvature corrections 
to Einstein-Maxwell-Gauss-Bonnet gravity that maintain second-order field equations with respect 
to the metric under conditions of spherical symmetry.  
We employ the action 
\begin{align}
I = \int d^{D} x \sqrt{-g} & \left( - 2 \Lambda + \mathcal{L}_1 + \frac{\lambda}{(D-3)(D-4)} \mathcal{L}_2 \right. \nonumber \\
 &+ \left. \frac{8\mu}{(D-3)(D-6)} \mathcal{L}_3 - F_{\mu \nu}F^{\mu \nu}   \right)
\label{action}
\end{align}
where the couplings have been redefined from their original formulation \cite{Myers2010,Myers2010a,Oliva2010} to
allow  a direct and straightforward comparison with Lovelock gravity.
Here  $D$ is the number of dimensions, 
$F_{\mu \nu} = \partial_{[\mu}A_{\nu]}$, $\mu$ and $\lambda$ are the correction terms' coefficients,
$\mathcal{L}_1 = R$ is the Ricci scalar, 
$\mathcal{L}_2 = R_{\mu \nu \gamma \delta} R^{\mu \nu \gamma \delta} - 4 R_{\mu \nu} R^{\mu \nu} + R^2$
is the Gauss-Bonnet Lagrangian, and $\mathcal{L}_3$ is the $3^{rd}$ order quasitopological gravity term.
This term has the form
\begin{align}
\mathcal{L}_3 &= \frac{2D-3}{3D^2 - 15D + 16} \left[ \vphantom{\left( \frac{3(3D - 8)}{8} R_{\mu \alpha \nu \beta} R^{\mu \alpha \nu \beta} R \right.} {{{R_\mu}^\nu}_\alpha}^\beta {{{R_\nu}^\tau}_\beta}^\sigma {{{R_\tau}^\mu}_\sigma}^\alpha \right.  \nonumber \\
              & + \frac{1}{(2D - 3)(D - 4)} \left( \frac{3(3D - 8)}{8} R_{\mu \alpha \nu \beta} R^{\mu \alpha \nu \beta} R \right. \nonumber \\
              &  - 3(D-2) R_{\mu \alpha \nu \beta} {R^{\mu \alpha \nu}}_\tau R^{\beta \tau} + 3D\cdot R_{\mu \alpha \nu \beta} R^{\mu \nu} R^{\alpha \beta} \nonumber \\
              & \left. + 6(D-2) {R_\mu}^\alpha {R_\alpha}^\nu {R_\nu}^\mu \right. \nonumber \\
              & \left. \left.- \frac{3(3D-4)}{2} {R_\mu}^\alpha {R_\alpha}^\mu R + \frac{3D}{8} R^3 \right) \right]
\label{quasitop}
\end{align}
and is only effective in higher dimensions ($D>4$), while becoming a surface term
in 6 dimensions \cite{Myers2010}.

 Of particular importance to this work is the scaling of the higher-curvature coupling coefficients.
Written in this form, $\lambda$ and $\mu$ are dimensionful, since the Riemann terms obey scaling $\sim L^{-2}$,
where $L$ is some lengthscale set by the metric coordinates. Therefore, $\lambda \sim L^{2}$ and $\mu \sim L^4$.
Later, when we use an Eulerian argument to construct an extended Smarr formula from a $1^{st}$ law, the scaling of these terms will set the coefficients in the Smarr formula.

In the remainder of this work, we will first generalize the exact solution found in \cite{Brenna2012a}
to both $D=5$ as well as $D=7$.
We will compare the resulting thermodynamic behaviour of the $D=7$ solution with previous work from 
cubic Lovelock gravity \cite{Frassino2014}, and we will discuss any new features that arise as we take $D=5$.

\section{Exact Solution}

The metric ansatz we employ is
\begin{equation}\label{metric}
ds^2 = -\frac{r^2}{L^2} f(r) dt^2 + \frac{L^2 dr^2}{r^2 g(r)} + r^2 d\Omega_k^2
\end{equation}
where $d\Omega_k^2$ is the metric of a constant curvature hypersurface
\begin{align}
d\Omega_k^2 &=d{\theta_1}^2 + \\
& \frac{\sin^2 {\left(\sqrt{k} \theta_1\right)}}{k} \left( d{\theta_2}^2 + \displaystyle\sum\limits_{i=3}^{D-2} \displaystyle\prod\limits_{j=2}^{i-1} \sin^2{\theta_j } d\theta_i^2 \right) \nonumber
\end{align}
The parameter $k=-1,0,1$, corresponds to hyperbolic, flat, and spherical geometries, respectively. Topological black holes are obtained by putting appropriate identifications on the former two \cite{Mann1997}.
We parameterize the Maxwell gauge field with a function $h(r)$ as
\begin{equation}\label{gfield}
A_t = q \frac{r}{L} h(r) .
\end{equation}
Substituting this ansatz into the action, using 
\begin{equation}
\Lambda =-\frac{(D-1)(D-2)}{2 L^2}
\end{equation}
as the cosmological constant and simplifying using integration by parts to remove terms proportional to derivatives of 
$f(r)$, we find
\begin{widetext}
\begin{equation}\label{simpleAction}
I = \frac{1}{L^2}\int d^{D-1} x \int dr  \sqrt{\frac{f}{g}} \left(\left\{(D-2)r^{D-1} \left( 1-\kappa + \frac{\lambda}{L^2} \kappa^2 - \frac{\mu}{L^4} \kappa^3 \right) \right\}' +2q^2r^{D-2}\frac{g}{f} \left[(rh)' \right]^2  \right)
\end{equation}
\end{widetext}
where a prime denotes differentiation with respect to $r$ and $\kappa = g(r) - L^2 k / r^2$.
Note that in eq.~(\ref{simpleAction}) the determinant of the angular part of the metric has been suppressed, since it is unimportant in obtaining the field equations.  

We substitute $f(r) = N^2(r)g(r)$ and vary the action with respect to  $g(r)$, $N(r)$ and $h(r)$ to obtain the field equations  
\begin{align}
\left(-1 + \frac{2 \lambda}{L^2} \kappa - \frac{3 \mu}{L^4} \kappa^2 \right)N' &= 0
\label{fieldEquations1}
\\
\left[(D-2)r^{D-1} \left(1- \kappa + \frac{\lambda}{L^2} \kappa^2 - \frac{\mu}{ L^4} \kappa^3 \right) \right]' & \nonumber \\
= \frac{2q^2r^{D-2}}{N^2}\left[(rh)'\right]^2  &
\label{fieldEquations2}
\\
\left(\frac{r^{D-2}\left(rh\right)'}{N}\right)' &=0.
\label{fieldEquations3}
\end{align}

Clearly, $N= 1$ solves equation (\ref{fieldEquations1}) and this solution will be taken for the remainder of the paper. 
Finding $h(r)$ from equation (\ref{fieldEquations3}) and solving for the gauge field from (\ref{gfield}), we find
\begin{equation}
A_{t} = \sqrt{\frac{(D-2)}{2(D-3)}}\frac{q}{r^{D-3}}.
\end{equation}
Solving the remaining differential equation, we obtain the fundamental relationship
\begin{equation}\label{masterEquation}
1 - \tilde{\kappa} + \frac{\lambda}{L^2} \tilde{\kappa}^2 - \frac{\mu}{ L^4} \tilde{\kappa}^3 = \frac{L^2 m}{r^{D-1}} -  \frac{q^2 L^2}{r^{2(D-2)}}
\end{equation}
where
\begin{equation}
\tilde{\kappa} = \frac{L^2}{r^2}\left(F(r) - k \right)
\end{equation}
and $F(r) = r^2g(r)/L^2$ is defined such that the metric reads: $-F(r) dt^2 + \cdots $.   Note that if we were to use the function $g(r)$ as defined previously, we would be excluding factors of $L$ from the mass and charge.  Since in extended phase space we consider $L$ to be a parameter that is varied, it is more convenient to  write expressions in terms of $F(r)$.
Note also that the mass parameter $m$ and the charge parameter $q$ have been defined such that we recover the RN-AdS solution in the form presented in~\cite{Gunasekaran2012} when we take $\mu=\lambda=0$ .

\section{Smarr Relation and $1^{st}$ Law}

When considering the thermodynamics of these black holes it is useful to work explicitly with eq.~(\ref{masterEquation}).  The utility of using this expression is that one is not restricted to considering only one branch of the solution. 
As a result of this, it is natural to expect that all thermodynamic quantities should limit to the standard $D-$dimensional RN-AdS results as $\mu = \lambda = 0$.

We now move on to considering the thermodynamics for these black holes in extended phase space, where all dimensionful parameters are considered to be thermodynamic variables \cite{Kastor:2010gq}. We employ the extended first law,
\begin{equation}
dM  = TdS + VdP +\Phi dQ+ \Psi_{\lambda}d\lambda +  \Psi_{\mu} d\mu 
\end{equation}
where 
\begin{align}
M &= \frac{(D-2)}{16 \pi} \omega_{D-2} \, m 
\\
Q &= \frac{\sqrt{2(D-2)(D-3)}}{8 \pi} \omega_{D-2} \, q
\end{align}
as in~\cite{Gunasekaran2012,Kastor:2010gq}, where
\begin{equation}
\omega_{D} \equiv \frac{2 \pi^{(D+1)/2}}{\Gamma\left(\frac{D+1}{2}\right)}
\end{equation} is the surface area of a $D$-dimensional sphere.

 The potentials introduced by the quasitopological and Gauss-Bonnet parameters follow from the procedure
introduced in \cite{Frassino2014}. The entropy  of this solution was found to be
\begin{equation}\label{entropyD}
S = \frac{r_h^{D-2} \omega_{D-2}}{4 } \left( 1 + \frac{2 (D-2) \lambda k}{(D-4)r_h^2} + \frac{ 3(D-2) \mu k^2}{(D-6) r_h^4} \right)
\end{equation}
and the temperature can be found from
\begin{equation*}
T = \left. \left( \frac{\partial F(r) / \partial r}{ 4 \pi} \right) \right|_{r = r_h}
\end{equation*}
where $r_h$ is the outermost root of $F(r) = 0$. In practice, to compute $\partial F(r) / \partial r$ it is easiest to differentiate eq.~(\ref{masterEquation}) and solve for the derivative term. Doing so yields
\ba
T = && \frac{1}{4 \pi \left(r_h^4 + 2k \lambda r_h^2 + 3k^2\mu \right)}
\left[\frac{(D-1)r_h^5}{L^2} + (D-3)kr_h^3 \right. \nonumber\\
&&\left. +(D-5) k^2 \lambda r_h + \frac{(D-7)k^3 \mu }{r_h} - \frac{(D-3)Q^2}{r_h^{2D-9}}\right]
\ea

The remaining potentials can be calculated by requiring consistency of the first law.  Defining the pressure as
\begin{equation}
P = -\frac{\Lambda}{8 \pi} = \frac{(D-1)(D-2)}{16 \pi L^2} ,
\end{equation}
a straightforward calculation yields
\begin{equation}\label{Volume}
V = \frac{\omega_{D-2}}{D-1}r_h^{D-1}
\end{equation}
for the thermodynamic volume and 
\begin{align}
\Psi_{\lambda} &= \frac{\omega_{D-2} (D-2)}{16 \pi } k r_h^{D-5} \left(  k - \frac{8 \pi r_h T }{D-4} \right),
\\
\Psi_{\mu} &= \frac{\omega_{D-2} (D-2) }{16 \pi } k^2 r_h^{D-7} \left( k - \frac{12 \pi r_h T}{D-6} \right).
\end{align}
for the potentials conjugate to the couplings. One finds that these thermodynamic quantities satisfy the Smarr relation 
\begin{align*}
\left(D - 3 \right) M  = & \left( D - 2 \right) T S - 2 P V + \left( D - 3 \right) \Phi Q \\
& + 2 \lambda \Psi_{\lambda} + 4 \mu \Psi_{\mu}
\end{align*}
which is consistent with Eulerian scaling.

We are now in a position to examine these black hole solutions for critical behaviour in extended phase space. In the following sections we will adopt an approach which is similar  to that used in~\cite{Frassino2014} insofar as we will employ dimensionless thermodynamic parameters to simplify our investigation.     However (unlike the analysis performed in \cite{Frassino2014}) we will not restrict ourselves to  positive values of the coupling $\mu$ but will instead  explore the entire parameter space that is consistent with the entropy, pressure, and stability constraints  (or in other words a ghost-free vacuum) 
discussed in the following sections. The case of positive quasitopological coupling will be considered in section~\ref{positiveMuSec}, while section~\ref{negativeMuSec} will focus on negative quasitopological coupling.

\section{Positive $\mu$ thermodynamics}\label{positiveMuSec}

Analogous to~\cite{Frassino2014}, we define the dimensionless thermodynamic quantities\footnote{Note that the definition of the dimensionless charge differs from that in~\cite{Frassino2014}, while all other dimensionless quantities are the same.  This adjustment was made to the charge to account for the differences in coupling to the electromagnetic field.}
\begin{align}\label{dimensionlessRatios}
r_h &= v \mu^{1/4}, \;\;\;\;\;  \tq = \frac{q}{2} \sqrt{\frac{(D-2)(D-3)}{\pi}} \mu^{-(D-3)/4},
\nn\\
 &t= (D-2) \mu^{1/4} T.
\end{align}
where we have performed a rescaling of the charge to make comparisons with~\cite{Frassino2014} more convenient.  Note that from (\ref{Volume}) $r_h$ will be non-linearly related to $V$. It is more convenient to work with the specific volume $v$, than $V$.  Qualitatively this has no effect on the results we obtain, though care must be taken upon integration
over $v$ (for example in computations involving the equal-area law) since the measure of integration will contribute.

In terms of these quantities, along with the parameters
\begin{equation}\label{couplingRatio}
\alpha = \frac{\lambda}{\sqrt{\mu}} \;\;\;\;\;\; p  = 4\sqrt{\mu}P ,
\end{equation}
we can write the dimensionless equation of state in the form
\begin{align}\label{dimensionlessEOS}
p &= \frac{t}{v} - \frac{(D-2)(D-3)k}{4 \pi v^2} + \frac{2 k t \alpha }{v^3} -\frac{(D-2)(D-5) k^2 \alpha}{4 \pi v^4} 
\nn\\
&+ \frac{3 k^2 t}{ v^5} - \frac{(D-2)(D-7) k}{ 4 \pi v^6} + \frac{\tq^2}{v^{2(D-2)}}.
\end{align}
We see that eq.~(\ref{dimensionlessEOS}) is identical to the equation of state found in the study of $3^{rd}$ order Lovelock gravity (c.f. eq. (4.7) from \cite{Frassino2014}). Hence the same thermodynamic behaviour seen previously for $D \ge 7$ -- particularly multiple re-entrant phase transitions \cite{Frassino2014} and isolated critical points \cite{Dolan:2014vba} -- 
will be seen here as well.  

In this case we are permitted to study the $D=5$ behaviour, since quasitopological gravity is well-defined in five spacetime dimensions.  For $D=5$, some terms in (\ref{dimensionlessEOS}) vanish and others reverse in sign, and so we must directly analyze this particular case.  We shall carry out this analysis in the next subsection.

Next we compute the Gibbs free energy, $G = M-TS$ for this solution (in dimensionless form):
\begin{equation}
g = \frac{\mu^{(3-D)/4}}{\omega_{D-2}}G.
\end{equation}
Explicitly in the case $k= \pm 1$, this is given by 
\begin{widetext}
\begin{align}
g &= -\frac{1}{16 \pi \left( v^4 + 2\alpha k v^2 +3 \right)} \left[- \frac{3 (D-2) k v^{D-7}}{(D-6)} - \frac{3(D-2)(D-8) \alpha v^{D-5}}{(D-6)(D-4)}+\frac{4(D+3) k v^{D-3}}{D-6} - \frac{2(D-2) \alpha^2 k v^{D-3}}{D-4} \right. \nn\\
&\left. -\frac{\alpha (D-8) v^{D-1}}{D-4} + \frac{60 \pi p v^{D-1}}{(D-1)(D-6)} - k v^{D+1} + \frac{24 \pi \alpha k p v^{D+1}}{(D-1)(D-4)} + \frac{4 \pi p v^{D+3}}{(D-1)(D-2)} \right]
\nn\\
&+ \frac{\tq^2}{4(D-3)\left(v^4 + 2\alpha k v^2 +3 \right)v^{D-3}}\left[\frac{(2D-5) v^4}{D-2} + \frac{2(2D-7)\alpha k v^2}{D-4}  + \frac{3(2D-9)}{(D-6)}\right].
\label{dimlessgibbs}
\end{align}
\end{widetext}
Direct comparison between eq.~(\ref{dimlessgibbs}) and eq.~(4.9) in~\cite{Frassino2014} reveals that the two expressions are identical.  That is, the dimensionless Gibbs energy in quasitopological gravity is the same as the dimensionless Gibbs energy for $3^{rd}$ order Lovelock gravity.

We see from eq. (\ref{entropyD}) that for certain choices of parameters the entropy can be negative. 
In the thermodynamic considerations that follow, we consider only black holes with positive entropy to be physical.  For this reason, it is important to clarify the restrictions that determine the positivity of entropy.  The result is trivial in the $k=0$ case,  as the entropy is always positive.  In the cases $k=\pm 1$  the condition will depend on the number of spacetime dimensions.  Substituting the dimensionless volume $v$ from eq.~(\ref{dimensionlessRatios}) into the expression (\ref{entropyD}) for the entropy and solving for the zeros gives
\begin{equation}\label{positiveEntropyRequirement}
v_{\pm} = \sqrt{\left(\frac{D-2}{D-4}\right) \left[-k\alpha \pm \sqrt{\alpha^2 - \frac{3(D-4)^2}{(D-2)(D-6)}}  \right]}
\end{equation}
for the roots with $v>0$.  
For $D> 6$, the result is the same as for $3^{rd}$ order Lovelock gravity \cite{Frassino2014}: for
$k\alpha > 0$ the entropy is always positive, whereas  for $k\alpha < 0$, the entropy is positive provided $|\alpha| < \sqrt{\frac{3(D-4)^2}{(D-2)(D-6)}}$.
In the case $D=5$ positivity of entropy is satisfied for $v > v_+$,  regardless of the sign of $k\alpha$.

The requirement that asymptotically AdS regions exist for all branches of the solution was discussed in~\cite{Frassino2014}.  
We examine the asymptotic behaviour of the discriminant of eq.~(\ref{masterEquation}).
If the discriminant is asymptotically (at large $r$) greater than zero, all three
possible branches have real-valued solutions. 
However, if the discriminant is less than zero, only one of the branches will take
real values at large $r$. This corresponds to the case where only
the branch that does not reduce to either of the Gauss-Bonnet solutions as $\alpha_3 \rightarrow 0$
has AdS asymptotics.

\begin{figure}
\centering
\includegraphics[width=\linewidth]{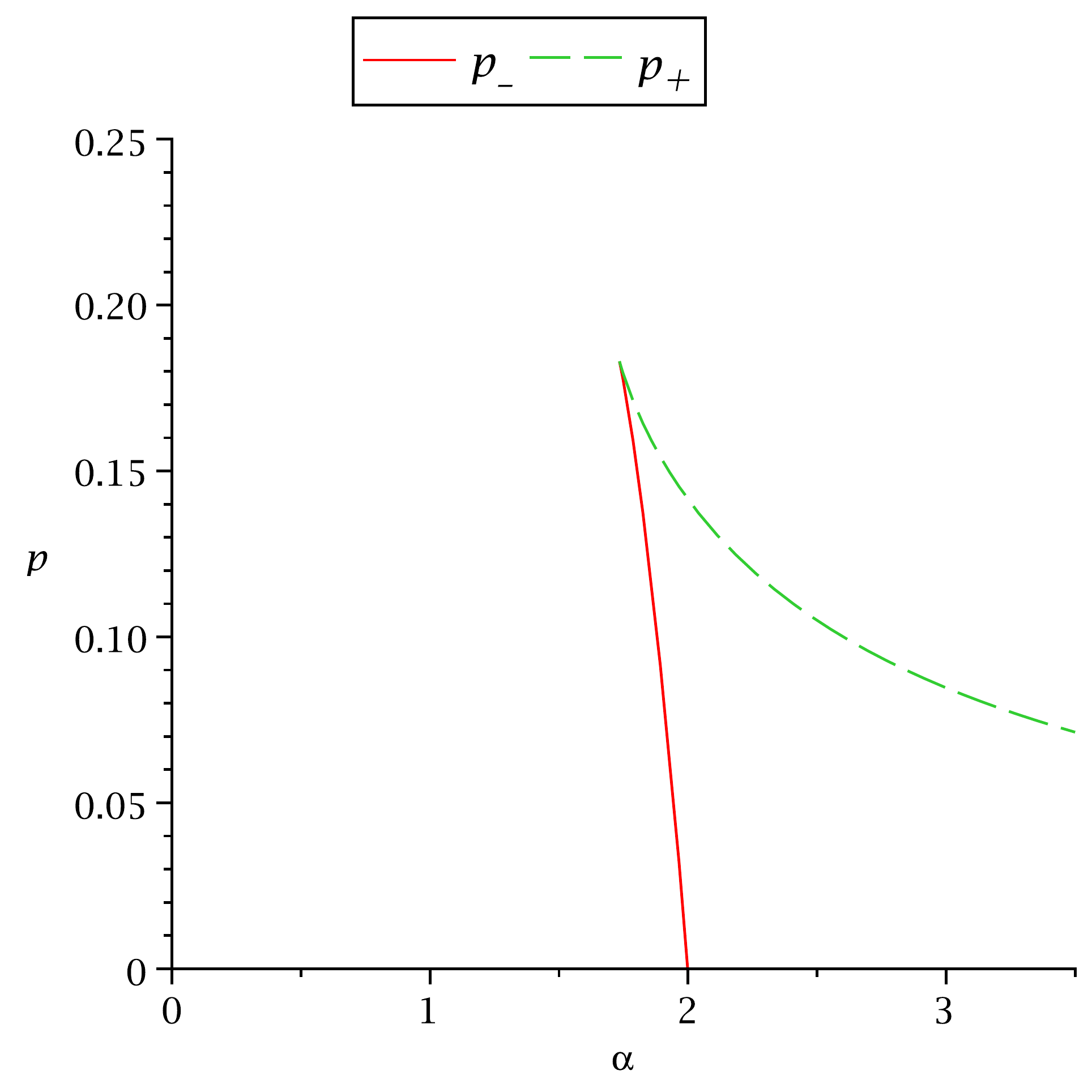}
\caption{ A plot of the regions bounded by $p_+$ and $p_-$ in 5 dimensions. Outside of the region,
in the majority of the plot, we have only one branch which has a valid AdS limit,
while inside the region all three branches have valid AdS limits.
}
\label{fig:regionplot}
\end{figure}

Taking the asymptotic limit, the equation defining $\tilde{\kappa}$ becomes
\begin{equation}\label{hinfeq}
h(\tilde{\kappa}_\infty) =  1 - \tilde{\kappa}_\infty + \frac{\lambda}{L^2} \tilde{\kappa}_\infty^2 - \frac{\mu}{ L^4} \tilde{\kappa}_\infty^3 
= 0
\end{equation}
and we find that its discriminant yields the inequality
\begin{equation}
\frac{18 \mu \lambda}{ L^6} - \frac{4 \lambda^3}{L^6} + \frac{\lambda^2}{L^4} - \frac{4 \mu}{L^4} - \frac{27 \mu^2}{ L^8} \ge 0
\end{equation}
which is quadratic in $1/L^2$ after multiplying through by $L^4$.  
Writing $L^2 = (D-1)(D-2)/16 \pi P$ and substituting 
the dimensionless parameters $p$ and $\alpha$ in eq.~(\ref{couplingRatio})
we obtain a discriminant of
\begin{align}
\Delta = \frac{\pi^2 p^2}{(D-2)^2(D-1)^2} &\left( 16 \alpha^2 - 64  \right) \nonumber \\
+ \frac{\pi^3 p^3}{(D-2)^3(D-1)^3}& \left( 1152 \alpha - 256 \alpha^3 \right) \nonumber \\
- \frac{6912 \pi^4 p^4}{(D-2)^4(D-1)^4} &
\end{align}
which equals zero when
\begin{equation}
p_{\pm} = \frac{(D-1)(D-2)}{108 \pi} \left[ 9 \alpha - 2\alpha^3 \pm 2 \left(\alpha^2 - 3\right)^{3/2}\right]
\end{equation}
denoting a transition in the number of solutions with AdS limits.
This is exactly the same as for the cubic Lovelock case, but with the important distinction that we can now examine the
behaviour in 5 dimensions, as seen in Figure \ref{fig:regionplot}.

We can examine the stability of the AdS vacuum
to show that all regions of Figure \ref{fig:regionplot} possess at least one branch with a
ghost-free AdS vacuum.
We do this by applying the condition \cite{Myers2010}  
\be 
h(\tilde{\kappa}_\infty) =  1 - \tilde{\kappa}_\infty + \frac{\lambda}{L^2} \tilde{\kappa}_\infty^2 - \frac{\mu}{ L^4} \tilde{\kappa}_\infty^3 
\ee 
on the ghosty graviton, which will yield $h'(\tilde{\kappa}_\infty) > 0$ if the graviton is a ghost.
This is because $h'(\tilde{\kappa}_{\infty})$ appears as the prefactor
to the standard equations of motion for perturbation about the AdS vacuum,
and determines the sign of the kinetic term in the propagator for the graviton.
Substituting $\tilde{\kappa}$ as a solution of
\begin{equation*}
0 = 1 - \tilde{\kappa}_\infty + \frac{\lambda}{L^2} \tilde{\kappa}_\infty^2 - \frac{\mu}{ L^4} \tilde{\kappa}_\infty^3
\end{equation*}
which is the condition for AdS asymptotics, we find that $h'(\tilde{\kappa}_\infty) < 0$ for two of the three branches
when both the conditions $p>p_-$ and $p<p_+$ are satisfied (i.e. this is the region where three AdS limits exist). 
Furthermore, in the region with only one valid AdS limit, $h'(\tilde{\kappa}_\infty) < 0$ and so we again have a solution
with a ghost-free vacuum.

Another important consideration for the thermodynamics of these black holes is the existence of a thermodynamic singularity.  It was shown in~\cite{Frassino2014} that such a singularity exists when $k=-1$ and when
\begin{equation}
\left. \frac{\partial p}{\partial t} \right|_{v=v_s} = 0
\end{equation}
for $D=7$ Lovelock gravity.  In view of the preceding analysis, the same situation holds in quasitopological gravity.  Evaluating the condition above for the case $k = \pm 1$, one finds that the real solution for the specific volume given by
\begin{equation}\label{vsing}
v_s = \sqrt{-k \alpha \pm \sqrt{\alpha^2 -3}}
\end{equation}
results in a thermodynamic singularity. 
It is clear that only  for $k\alpha < 0$  can a thermodynamic singularity occur. 
At this value of the specific volume the temperature diverges.  However, one can find a pressure for which the temperature is in fact well behaved at this singularity.  In~\cite{Frassino2014} this was done by solving
\begin{equation}
\left. \frac{\partial p}{\partial v} \right|_{t=t_s} = 0
\end{equation}
for the temperature,  inserting it and $v_s$ into the equation of state to obtain the pressure at this singular point, $p_s$.  

Since the result here is identical to that found in the case of $3^{rd}$ order Lovelock gravity, we see that cubic quasitopological
gravity has the same isolated critical points as $3^{rd}$-order Lovelock gravity.  These occur for $\alpha=\sqrt{3}$ (ensuring their entropy is positive), and 
correspond to the existence of two swallowtails, giving rise to two first-order phase transitions between small and large black holes. This phase transition is  characterized by non-standard critical exponents in the phase diagram \cite{Dolan:2014vba}.

\subsection{Thermodynamics in $D=5$}

In the previous section, we have shown that the extended phase space thermodynamics in $3^{rd}$ order quasitopological gravity is identical to the thermodynamics of $3^{rd}$ order Lovelock gravity, as considered in~\cite{Frassino2014}.  For this reason, we now study the results in $D=5$ to see if there exist any new thermodynamic features not present in the Lovelock case, and to see which of the thermodynamic results from Lovelock gravity are consistent with the constraints of $D=5$.

The first thing we note is that thermodynamic singularities in $D=5$ are not physical.  They correspond to black holes with negative entropy, since upon  comparison of eq.~(\ref{positiveEntropyRequirement}) with eq.~(\ref{vsing})  we find  $v_+ > v_s$. Hence cubic quasitopological gravity has no special isolated critical points in five dimensions. 

Moving on to a more detailed study, we note that due to the form of the equation of state, there is no interesting thermodynamic behaviour when $k=0$.  
Henceforth we examine only spherical and hyperbolic black holes.

\subsubsection{Spherical}

In the case $k=1$, the conditions for critical points,
\begin{equation}
\frac{\partial p}{\partial v} = 0, \quad \frac{\partial^2 p}{ \partial v^2} =0. 
\end{equation}
can be solved analytically, although the resulting expressions for $t_c, v_c$ and $p_c$ are not illuminating.  The result is that there exists one critical point which satisfies the constraint conditions provided $\alpha > \,\,\sim 1.9909$ (smaller values of $\alpha$ result in a $p_c$ that is smaller than $p_-$).  

We can calculate the critical exponents corresponding to this critical point by expanding the equation of state in terms of the parameters
\begin{equation}
\omega = \frac{v}{v_c} - 1, \;\;\; \tau = \frac{t}{t_c} - 1.
\end{equation}
In terms of these parameters our expansion takes the form,
\begin{equation}
\frac{p}{p_c} = 1 + A\tau + B \tau \omega + C\omega^3 + \mathcal{O}(\tau \omega^2, \omega^4)
\end{equation}
where $A,B$ and $C$ are complicated $\alpha$-dependent coefficients.  However, the important feature of this expansion is the form, which is identical to the cubic Lovelock case, and to the van der Waals fluid.
 Because the equation of state entirely governs the macroscopic behaviour of the system,
the critical exponents are therefore given by the standard mean field theory results:
\begin{equation}\label{critExps}
\alpha = 0, \;\;\; \beta = \frac{1}{2}, \;\;\; \gamma = 1, \;\;\; \delta = 3.
\end{equation}

\subsubsection{Hyperbolic}
We begin by mentioning that there are no isolated critical points found in $D=5$.  In the case of Lovelock gravity, isolated critical points were found in $D=7$ and higher and correspond to the case where the critical point occurs at the thermodynamic singularity.  However, as we discussed in the previous section, when $D=5$ and $k=-1$ the thermodynamic singularity occurs when the entropy is negative.  Thus the black holes exhibiting the isolated critical point are unphysical.  

In fact, the $D=5$ hyperbolic black holes appear to exhibit no critical behaviour at all.  When one solves for $v_c$ it is found that there are solutions for two critical points.  However, when one compares the solution for $v_c$ and $p_c$ against the constraints on pressure and entropy, it is found that the solutions violate either one or both of these constraints.  Thus, the critical behaviour for the hyperbolic black holes is unphysical.

One can consider the thermodynamics outside the pressure wedge of Figure \ref{fig:regionplot}, since there do exist legitimate black hole solutions here.  For example, in the case where $\alpha$ is negative with $k=-1$,  for small $p$ we see discontinuities
in the Gibbs free energy which could be studied further to perhaps
uncover new critical behaviour. However, because this region exists
beyond the pressure wedge of Figure \ref{fig:regionplot}, we cannot be certain that
the branch we are considering has valid AdS asymptotics; it may be that
this analysis jumps to branches for which $f(r)$ terminates at finite $r$.  
Ensuring that the asymptotics of the branches considered are well-behaved would require a non-trivial analysis of each branch, 
since the asymptotic structure will be a function of $m$, $\tq$, $\alpha$ and $p$.  We shall not consider this case further.

\section{Negative $\mu$ thermodynamics} \label{negativeMuSec} 

We now wish to consider the thermodynamic  behaviour of black holes which are solutions to the field equations \eqref{fieldEquations2} in the case of negative coupling.  We wish to employ the same calculational machinery that was used in the previous section, but in order for this to be consistent we must substitute everywhere in our thermodynamics $\mu \to - \mu$.  This substitution allows us to explore the regions of negative coupling uses the same dimensionless parameters as before. The fundamental relationship for solutions of the field equations upon substitution $\mu \to -\mu$ becomes,
\begin{equation}\label{masterEquationNegative}
1 - \tilde{\kappa} + \frac{\lambda}{L^2} \tilde{\kappa}^2 + \frac{\mu}{ L^4} \tilde{\kappa}^3 = \frac{L^2 m}{r^{D-1}} -  \frac{q^2 L^2}{r^{2(D-2)}}
\end{equation}
where, as before,
\begin{equation}
\tilde{\kappa} = \frac{L^2}{r^2}\left(F(r) - k \right)
\end{equation}
with $F(r) = r^2g(r)/L^2$ is defined so that the metric reads: $-F(r) dt^2 + \cdots $.  Similarly, any thermodynamic quantities can be obtained by substituting $\mu \to -\mu$ in the aforementioned results.  We can then employ the same dimensionless thermodynamic parameters as before
\begin{align}\label{dimensionlessRatios2}
r_h &= v \mu^{1/4}, \quad  \tq = \frac{q}{2} \sqrt{\frac{(D-2)(D-3)}{\pi}} \mu^{-(D-3)/4},
\nn\\
 &t= (D-2) \mu^{1/4} T, \quad \alpha = \lambda/\sqrt{\mu}.
\end{align} 

In terms of the quantities from eq. \eqref{dimensionlessRatios2}, the equation of state now reads,
\ba \label{eosNegative}
p &=& \frac{t}{v} - \frac{(D-2)(D-3) k}{4 \pi v^2} + \frac{2 \alpha k t}{v^3} - \frac{(D-2)(D-5) k^2 \alpha}{4\pi v^4} 
\nn\\
&-& \frac{3 k^2 t}{v^5} + \frac{(D-2)(D-7) k}{4 \pi v^6} + \frac{\tq^2 }{v^{2(D-2)}}
\ea
and the dimensionless Gibbs free energy is given by
\begin{widetext}
\begin{align}
g &= -\frac{1}{16 \pi \left( v^4 + 2\alpha k v^2 - 3 \right)} \left[- \frac{3 (D-2) k v^{D-7}}{(D-6)} + \frac{3(D-2)(D-8) \alpha v^{D-5}}{(D-6)(D-4)} - \frac{4(D+3) k v^{D-3}}{D-6} - \frac{2(D-2) \alpha^2 k v^{D-3}}{D-4} \right. \nn\\
&\left. -\frac{\alpha (D-8) v^{D-1}}{D-4} - \frac{60 \pi p v^{D-1}}{(D-1)(D-6)} - k v^{D+1} + \frac{24 \pi \alpha k p v^{D+1}}{(D-1)(D-4)} + \frac{4 \pi p v^{D+3}}{(D-1)(D-2)} \right]
\nn\\
&+ \frac{\tq^2}{4(D-3)\left(v^4 + 2\alpha k v^2 - 3 \right)v^{D-3}}\left[\frac{(2D-5) v^4}{D-2} + \frac{2(2D-7)\alpha k v^2}{D-4}  - \frac{3(2D-9)}{(D-6)}\right].
\label{dimlessgibbsNegative}
\end{align}
\end{widetext}

We once again restrict our attention to only those black holes which have positive entropy.  In this case, the zeroes of the entropy are
\begin{equation}\label{positiveEntropyRequirementNegative}
v_{\pm} = \sqrt{\left(\frac{D-2}{D-4}\right) \left[-k\alpha \pm \sqrt{\alpha^2 + \frac{3(D-4)^2}{(D-2)(D-6)}}  \right]}.
\end{equation}
The positivity of entropy  is straightforward for $D \ge 7$ since the entropy will be positive provided $ v > v_+$ for $k = \pm 1$.  The situation is more complicated in $D=5$ and is most simply captured by rewriting the bound from (\ref{positiveEntropyRequirementNegative}) as 
\be\label{posEntNegMu}
\alpha k >  -\frac{1}{6}\frac{v^4 + 9}{v^2}
\ee
where $v$ is the dimensionless specific volume defined earlier.  

We must also be aware of the asymptotic behaviour of the different branches of the solutions for this case.  Substituting the dimensionless parameters from eq. \eqref{dimensionlessRatios2} into eq.~\eqref{masterEquationNegative} and taking the asymptotic limit, we find that the discriminant is given by
\ba 
\Delta &=& \frac{16 \pi^2}{(D-1)^2(D-2)^2}\bigg [ -\frac{432 \pi^2 p^4}{(D-2)^2(D-1)^2} 
\nn\\
 &-& \left(\frac{16 \pi}{(D-1)(D-2)}  \alpha^3  +\frac{72 \pi }{(D-1)(D-2)} \alpha  \right) p^3 
 \nn\\
 &+& \left( 4 + \alpha^2 \right)p^2   \bigg ] 
\ea
and is equal to zero for
\be\label{presConstraintNeg} 
p_{\pm} = \frac{(D-1)(D-2)}{108 \pi} \left[- 9 \alpha - 2\alpha^3 \pm 2 \left(\alpha^2 + 3\right)^{3/2}\right].
\ee
From this we see that $p_-$ is strictly negative  (and so does not yield asymptotically AdS behaviour)
while on  the other hand, $p_+$ is strictly positive.  A simple analysis reveals that $\Delta > 0 $ for $p < p_+$ and $\Delta < 0$ for $p > p_+$.  From this we can conclude that in the region where $p > p_+$, eq. \eqref{masterEquationNegative} has one real branch and two complex conjugate branches, while in the region $0 < p < p_+$ all branches admit AdS asymptotics.  Along the line $p = p_+$ all branches have AdS asymptotics and two branches coincide. 
\begin{center}
\begin{figure}[htp]
\includegraphics[width=\linewidth]{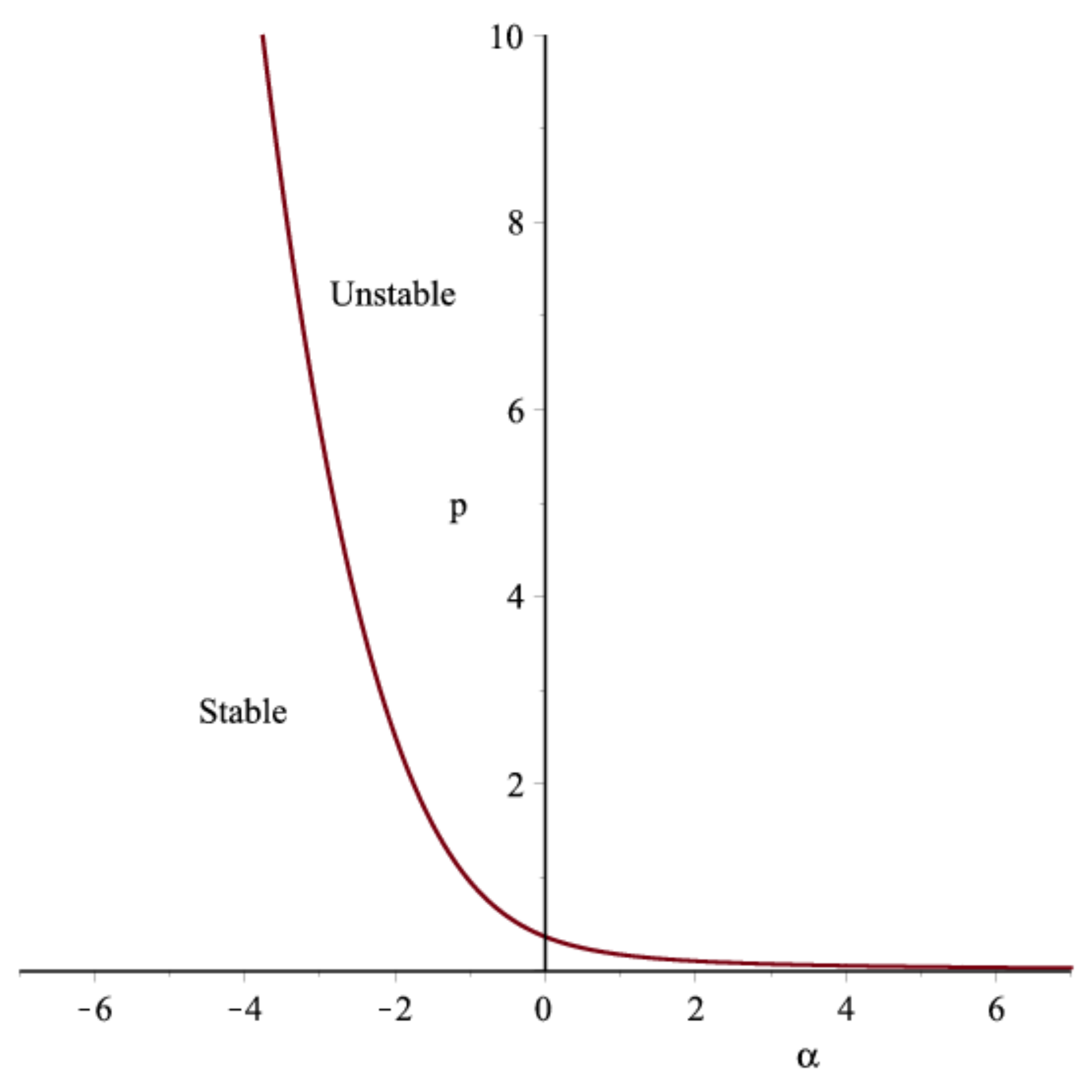}
\caption{A plot showing  the $p - \alpha$ parameter space in $D=5$ describing stability and existence of AdS vacua.  The dark red curve is a plot of $p_+$.  For $p> p_+$ only one branch admits AdS asymptotics, while for $0 < p \le p_+$ all branches admit AdS asymptotics.  Also shown is the stability of the AdS vacua, for $0 < p < p_+$ there is one stable branch. For all values of $D$ the plot is qualitatively the same, with the value of $D$ providing an overall scale for the plot.
}
\label{negMuAsymptotics}
\end{figure} 
\end{center}

As discussed in \cite{Myers2010,Myers2010a}, some AdS vacua are unstable in the sense that the graviton in these spacetimes may be a ghost.  One can determine whether or not this is the case by examining the slope of the asymptotic form of eq. \eqref{masterEquationNegative}
\be 
h(\tilde{\kappa}_\infty) =  1 - \tilde{\kappa}_\infty + \frac{\lambda}{L^2} \tilde{\kappa}_\infty^2 + \frac{\mu}{ L^4} \tilde{\kappa}_\infty^3 
\ee 
where $h'(\tilde{\kappa}_\infty)$ determines the sign of the kinetic term for the propagation of gravitons in the spacetime.  The kinetic term has the correct sign when  $h'(\tilde{\kappa}_\infty) < 0$ and the `wrong' sign when $h'(\tilde{\kappa}_\infty) > 0$. 
First, we write $h(\tilde{\kappa}_\infty)$ in terms of the dimensionless quantities, yielding
\be\label{hlarger} 
h(\tilde{\kappa}_\infty) = 1 - \tilde{\kappa}_\infty + \frac{4 \alpha \pi p \tilde{\kappa}_\infty^2 }{(D-1)(D-2)}  + \frac{16 \pi^2 p^2 \tilde{\kappa}_\infty ^3}{(D-1)^2(D-2)^2} 
\ee
and therefore,
\be\label{hprime}
h'(\tilde{\kappa}_\infty) = - 1  + \frac{8 \alpha \pi p \tilde{\kappa}_\infty }{(D-1)(D-2)}  + \frac{48 \pi^2 p^2 \tilde{\kappa}_\infty ^2}{(D-1)^2(D-2)^2}. 
\ee
Expressing  $\tilde{\kappa}_\infty$ in terms of $p$ and $\alpha$ by setting $h(\tilde{\kappa}_\infty) = 0$
we find that $h'(\tilde{\kappa}_\infty) > 0$ for $p \ge p_+$ for all branches, while there is one branch with $h'(\tilde{\kappa}_\infty) < 0$ for $p < p_+$.  Hence  the separation between stable and unstable regions coincides exactly with $p_+$.
 The existence of AdS asymptotics and the stability of these regions is captured graphically in Figure \ref{negMuAsymptotics}.  

\begin{center}
\begin{figure}[htp]
\includegraphics[width=\linewidth]{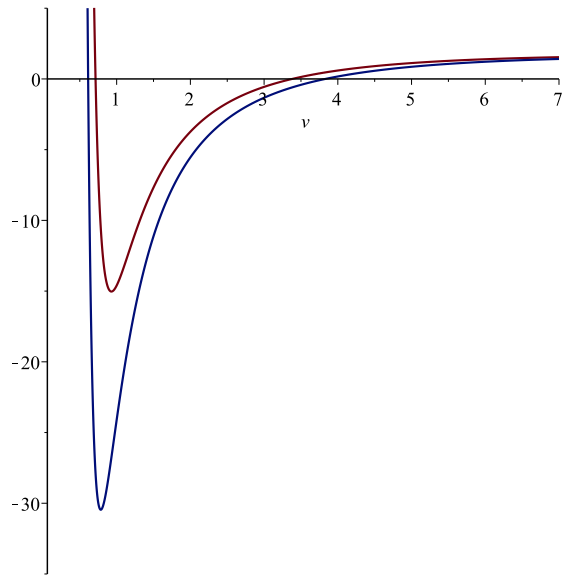}
\caption{A plot of the metric function $g(r)$ demonstrating black hole solutions for the stable branch with $p < p_+$.  The red curve is for $\mathfrak{m} = 25 $ and $\tq = 2$, the blue curve is for $\mathfrak{m} = 30 $ and $\tq = 2$.}
\label{blackHoleNegMu}
\end{figure} 
\end{center} 
 
Luckily for us, this stable AdS branch permits black hole solutions.  The easiest way to see this is to solve
\begin{equation}
1 - \kappa + \frac{\lambda}{L^2} \kappa^2 + \frac{\mu}{ L^4} \kappa^3 = \frac{L^2 m}{r^{D-1}} -  \frac{q^2 L^2}{r^{2(D-2)}}
\end{equation}
where
\begin{equation}
\kappa = g(r)-\frac{L^2}{r^2} k
\end{equation}
for $g(r)$ and note that a black hole solution would correspond to $g(r_h) = 0$. 
We use the dimensionless parameters eq. \eqref{dimensionlessRatios2} along with
\be
\mathfrak{m} = \mu^{-(d-3)/4}m 
\ee
as a rescaled mass parameter.  A representative plot of $g(r)$ is shown in Figure \ref{blackHoleNegMu}.  Before moving on to thermodynamic considerations we note that the analysis performed here indicates that we are free to consider the thermodynamics of these black holes in the case $\alpha < 0$, which corresponds to $\lambda, \mu < 0$.  This feature is captured in Figure \ref{negMuAsymptotics}, where we see that there exist stable AdS vacua in this case. Furthermore, since there are no stable vacua for $p \ge p_+$ we regard this region of parameter space as unphysical.

\subsection{Thermodynamic Singularity}

A general feature that emerges in the case of negative coupling is the existence of thermodynamic singularities 
for both $k= \pm 1$.  The $k=1$ result is novel since in the case of Lovelock gravity, thermodynamic singularities were only found in the case $k=-1$
 \cite{Frassino2014}. Here we outline some general considerations for the thermodynamic singularities that occur for negative coupling.  

As was discussed earlier, a thermodynamic singularity occurs when
\begin{equation}\label{tdsing}
\left. \frac{\partial p}{\partial t} \right|_{v=v_s} = 0 \; .
\end{equation}

Working with eq. \eqref{eosNegative} (first in the case where $\tq=0$), we find that eq. \eqref{tdsing} has the positive solution,
\be
v_{s_\pm} = \sqrt{-\alpha k \pm \sqrt{\alpha^2 + 3}}.
\ee
We immediately see that for $k=\pm 1$ only $v_{s_+}$ will be real, indicating only one thermodynamic singularity.  We can define the thermodynamic singular point $(p_s, v_s, t_s)$ by,
\ba
p_s &=& p_s(v_s, t_s) \quad 
\left. \frac{\partial p}{\partial t} \right|_{v=v_s} = 0, \quad 
\left. \frac{\partial p}{\partial v} \right|_{t=t_s} = 0.
\ea
which yields
\ba\label{tdsingpoint}
v_s &=& \sqrt{-\alpha k + \sqrt{\alpha^2 + 3}}
\nn\\
t_s &=& \frac{ D-2 }{12 \pi \alpha k v_s^3 + 2\pi v_s^5 - 30 \pi v_s }\bigg[ k(D-3) v_s^4 + 2 \alpha (D-5) v_s^2  
\nn\\
&+& \left. 3k (D-7) + \frac{4 \pi \tq^2}{v^{2(D-5)}} \right ]
\nn\\
p_s &=& \frac{(D-1)(D-2) k }{4 \pi \left(\alpha k - \sqrt{\alpha^2 + 3}  \right)^3\left( \alpha k \sqrt{\alpha^2+3} - \alpha^2 - 3\right)}
\nn\\
&\times & \bigg[ \alpha k (2\alpha^2 +  5) \sqrt{\alpha^2 + 3} - 2\alpha^4 - 8 \alpha^2 - 6) \bigg]
\nn\\
&+&\frac{\tq^2}{\left(\alpha k - \sqrt{\alpha^2 + 3} \right)^{D-2}}. 
\ea
Let us now make a few comments regarding the thermodynamic singularity.  Studying the expression for $p_s$ in eq. \eqref{tdsingpoint} in the case $k=-1$, we find  
\be 
p_s = p_+ + \frac{\tq^2}{\left( \alpha +  \sqrt{\alpha^2 + 3} \right)^{D-2}}
\ee
and so in the hyperbolic case the thermodynamic singularity occurs at $p_s \ge p_+$ and is therefore excluded by our pressure and stability constraints. 

\begin{center}
\begin{figure}[htp]
\includegraphics[width=\linewidth]{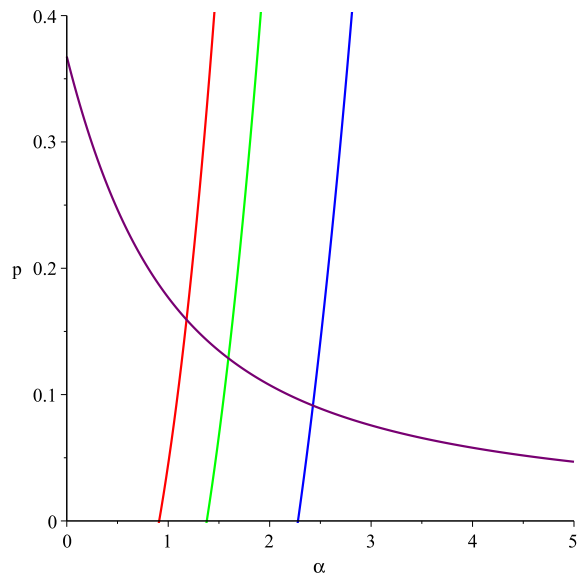}
\caption{A plot of $p_s$ for $k=1$ for $\tq = 0.8, 0.9, 1.0$ (blue, green, and red curves, respectively); the purple curve is $p_+$.  The thermodynamic singularity is physical for each value of $\tq$ in the region where $0 < p_s < p_+$.  The plot is for $D=5$ but the basic result is identical for $D \ge 7$ as well. } 
\label{tdsingPlot}
\end{figure} 
\end{center}

More interesting is the $k=1$ case, where we have,   
\be 
p_s = p_- + \frac{\tq^2}{\left(-\alpha +   \sqrt{\alpha^2 + 3} \right)^{D-2}}.
\ee
From eq. \eqref{presConstraintNeg}, $p_- < 0$, so when $\tq =0$ the thermodynamic singularity will 
occur for $\Lambda > 0$, and is therefore unphysical.  However it will always be possible to choose $\tq$ such that, for some range of $\alpha$, $p_s$ will occur for physically relevant pressures.  This is illustrated in Figure~(\ref{tdsingPlot}).  We will return to a deeper discussion of the thermodynamic singularity in the following sections.

\subsection{Thermodynamics in $D=5$}

In five dimensions the equation of state reads (for $k = \pm 1$),
\be
p = \frac{t}{v} - \frac{3 k }{2 \pi v^2} + \frac{2 k \alpha t}{v^3}  - \frac{3 t}{v^5} - \frac{3 k}{2\pi v^6} + \frac{\tq^2}{v^6}
\ee
and the conditions for a critical point reduce to
\ba\label{critPointsNegMu} 
t_c &=& \frac{3kv_c^4 + 9 k - 6 \pi  \tq^2}{\pi v_c \left(v_c^4 + 6 \alpha k v_c^2 - 15 \right)},
\nn\\
0 &=& -k v_c^8 + 6v_c^6 \alpha + (10\pi \tq^2 -60 k)v_c^4 
\nn\\
&+& (36\pi k \tq^2 - 54)\alpha v_c^2 - 30 \pi \tq^2 + 45 k.
\ea

\subsubsection{Spherical}

\begin{center}
\begin{figure}[htbp]
  \includegraphics[width=0.9\linewidth]{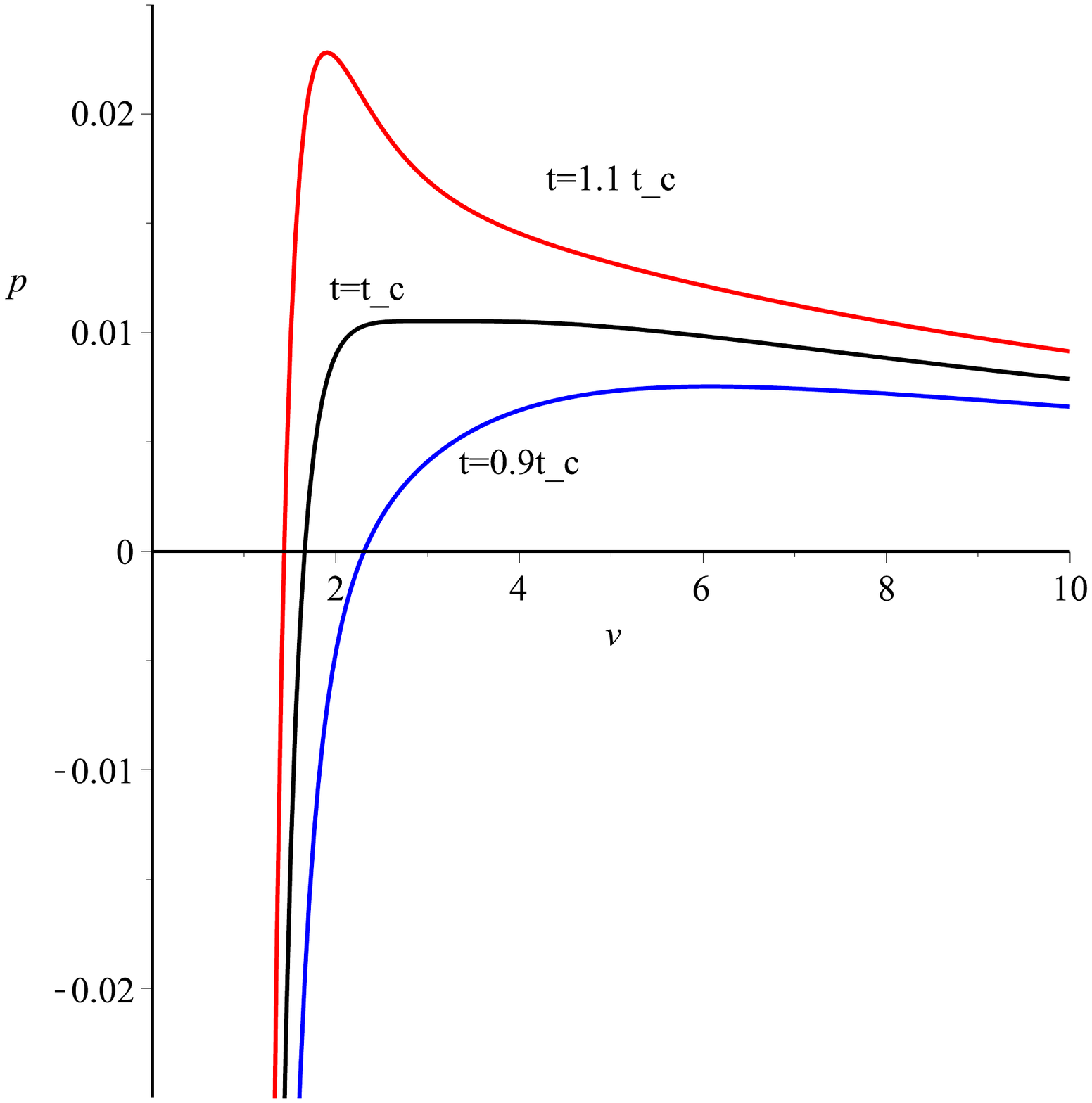}

  \includegraphics[width=0.9\linewidth]{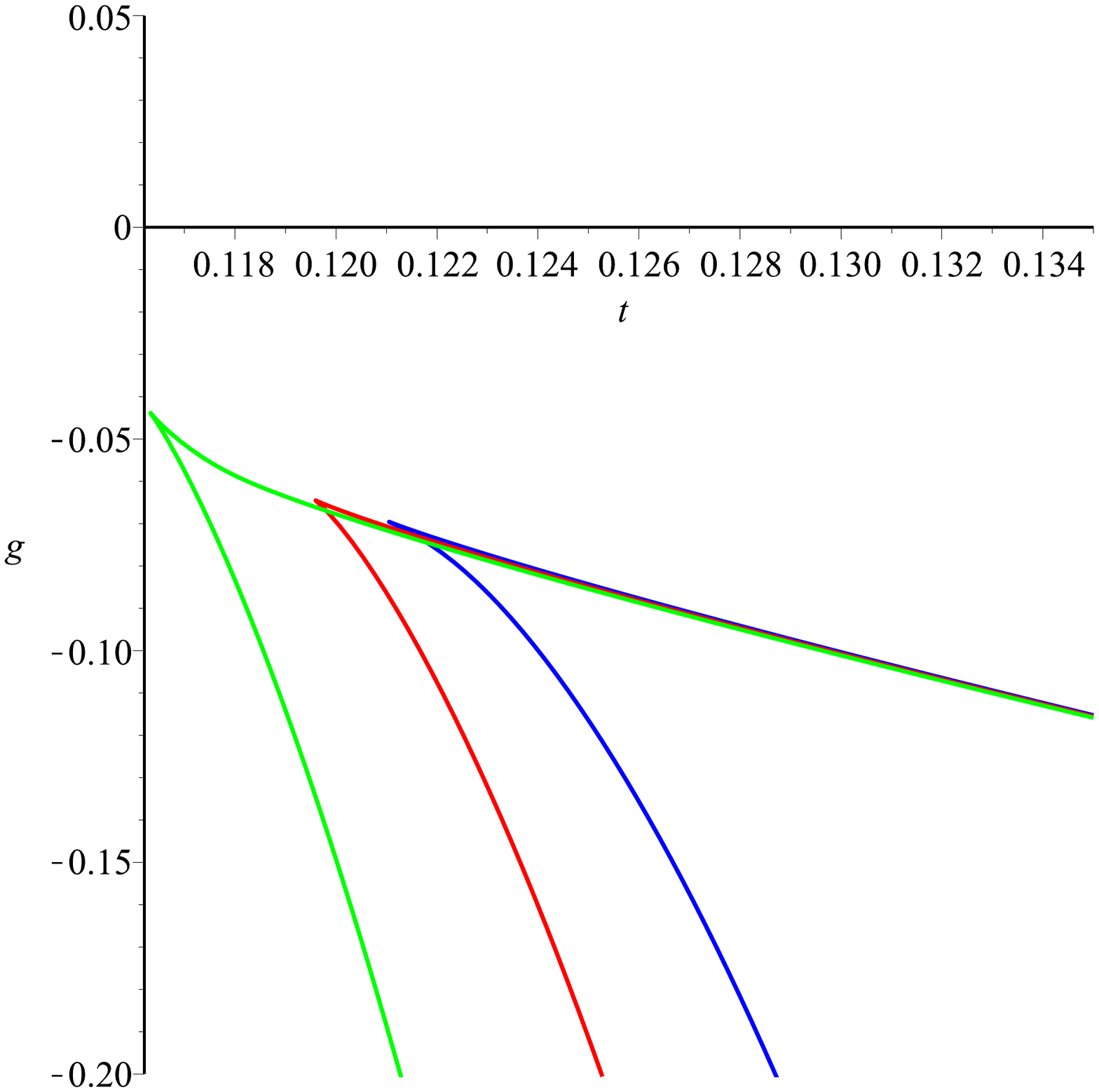}

\caption{\textit{Top}: $p-v$ plot for $\alpha = \sqrt{225 + 120\sqrt{15}}/9$ exhibiting an inflection point.  \textit{Bottom}: $g-t$ projection for $p=p_c, 1.1p_c$ and $0.9 p_c$ (red, blue, green, respectively).  The Gibbs free energy has a cusp, indicating that the inflection point is not a critical point.  }
\label{coalescedCPk1}
\end{figure}
\end{center}

We specialize first to the case of spherical horizons (i.e. $k=+1$) with $\tq=0$.   We find in this case that the number of possible critical points consistent with the constraints for a stable, asymptotically AdS vacuum breaks down as: 
\begin{eqnarray}
\alpha &<& \frac{\sqrt{225 + 120\sqrt{15}}}{9}  \quad   \textrm{\small zero possible critical points} \nonumber \\
\alpha &=& \frac{\sqrt{225 + 120\sqrt{15}}}{9} \quad   \textrm{\small single inflection point}  \nonumber\\
\alpha &\in& \left(\frac{\sqrt{225 + 120\sqrt{15}}}{9}, \frac{\sqrt{282 +102 \sqrt{17}}}{6}\right) \quad   
{{\textrm{\small two possible}}\atop{\textrm{\small critical points}}} \nonumber\\
\alpha &>& \frac{\sqrt{282 +102 \sqrt{17}}}{6}  \quad   \textrm{\small one possible critical point.} 
\end{eqnarray}
At the point $\alpha = \sqrt{225 + 120\sqrt{15}}/9$ the two inflection points `coalesce' into a single inflection point.  
This inflection point 
is not a physical critical point:  the Gibbs energy has a cusp at this combination of $(p,t,v)$, 
as shown in Figure~(\ref{coalescedCPk1}), and therefore no phase transition occurs.

\begin{center}
\begin{figure*}[htbp]
\centering
  \includegraphics[width=.35\linewidth]{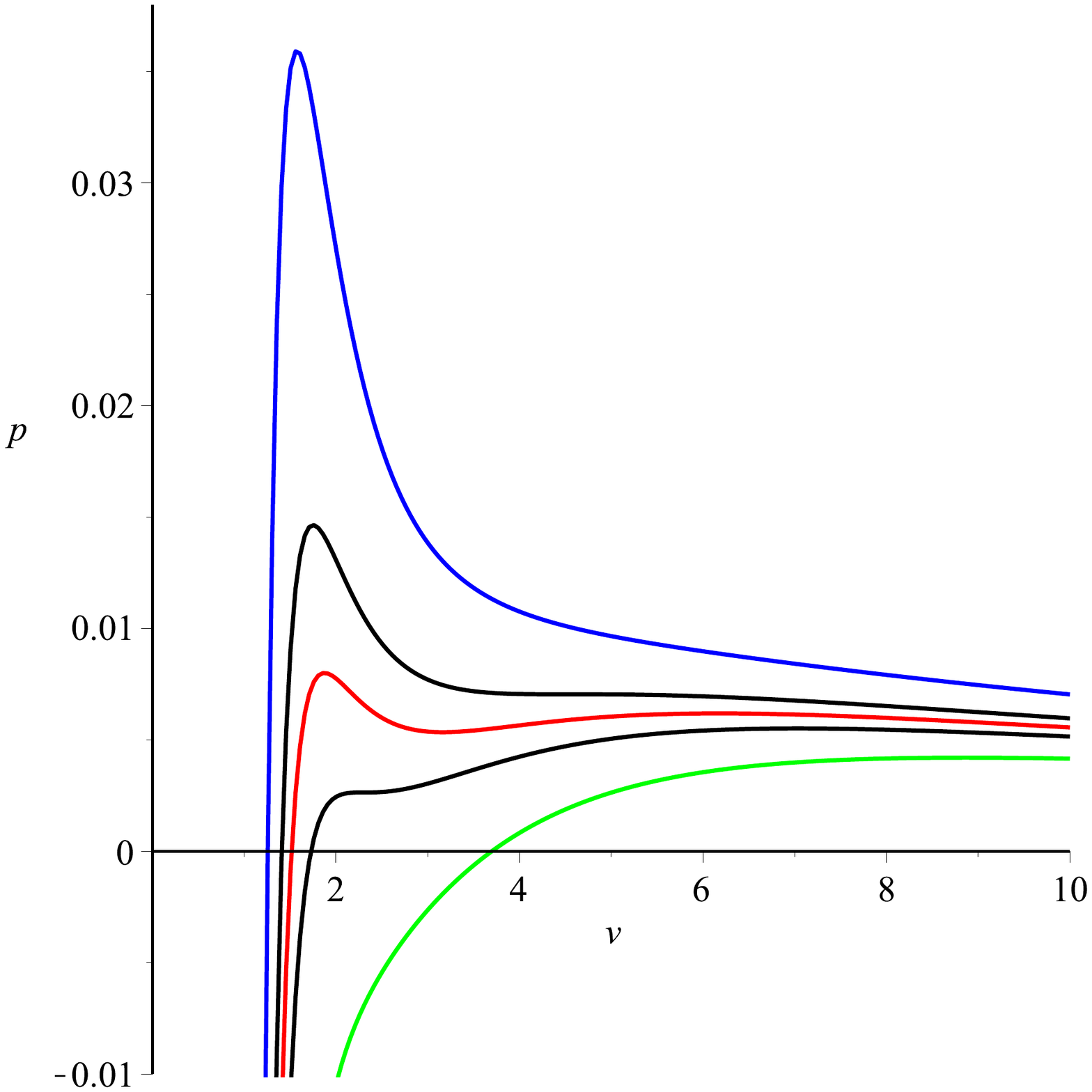}
\quad 
  \includegraphics[width=.35\linewidth]{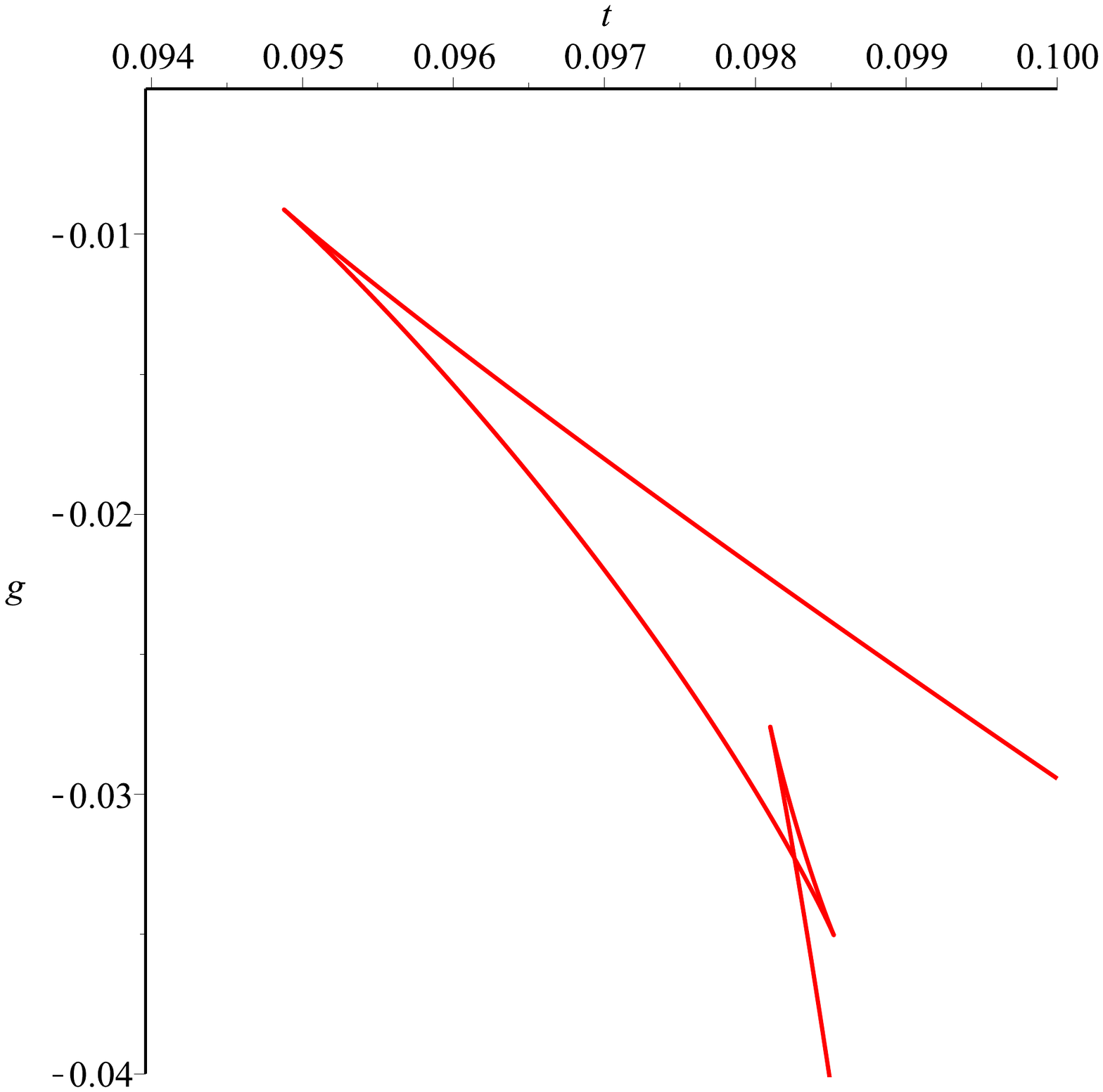}
\includegraphics[width=.35\linewidth]{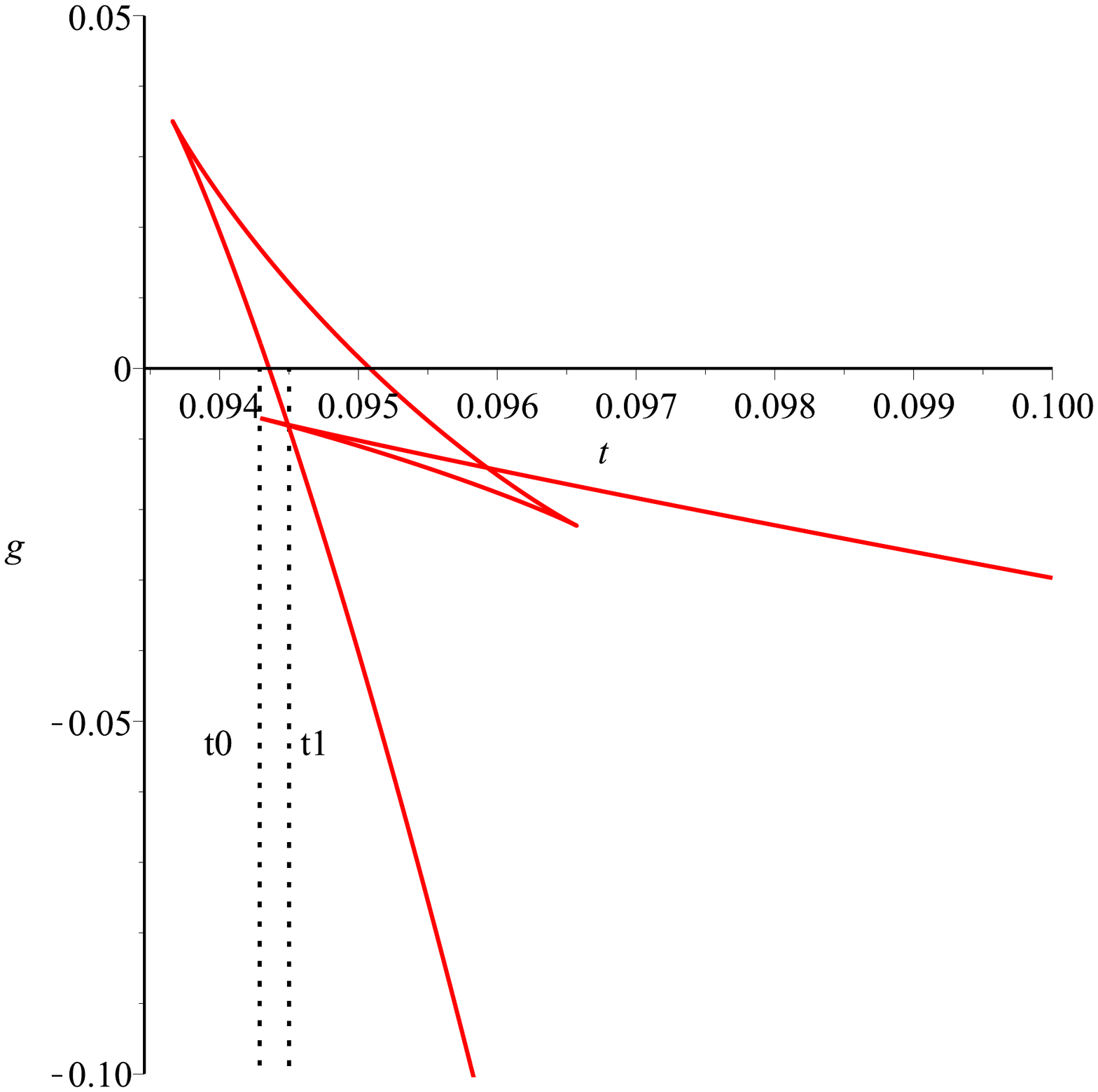}
\quad
  \includegraphics[width=.35\linewidth]{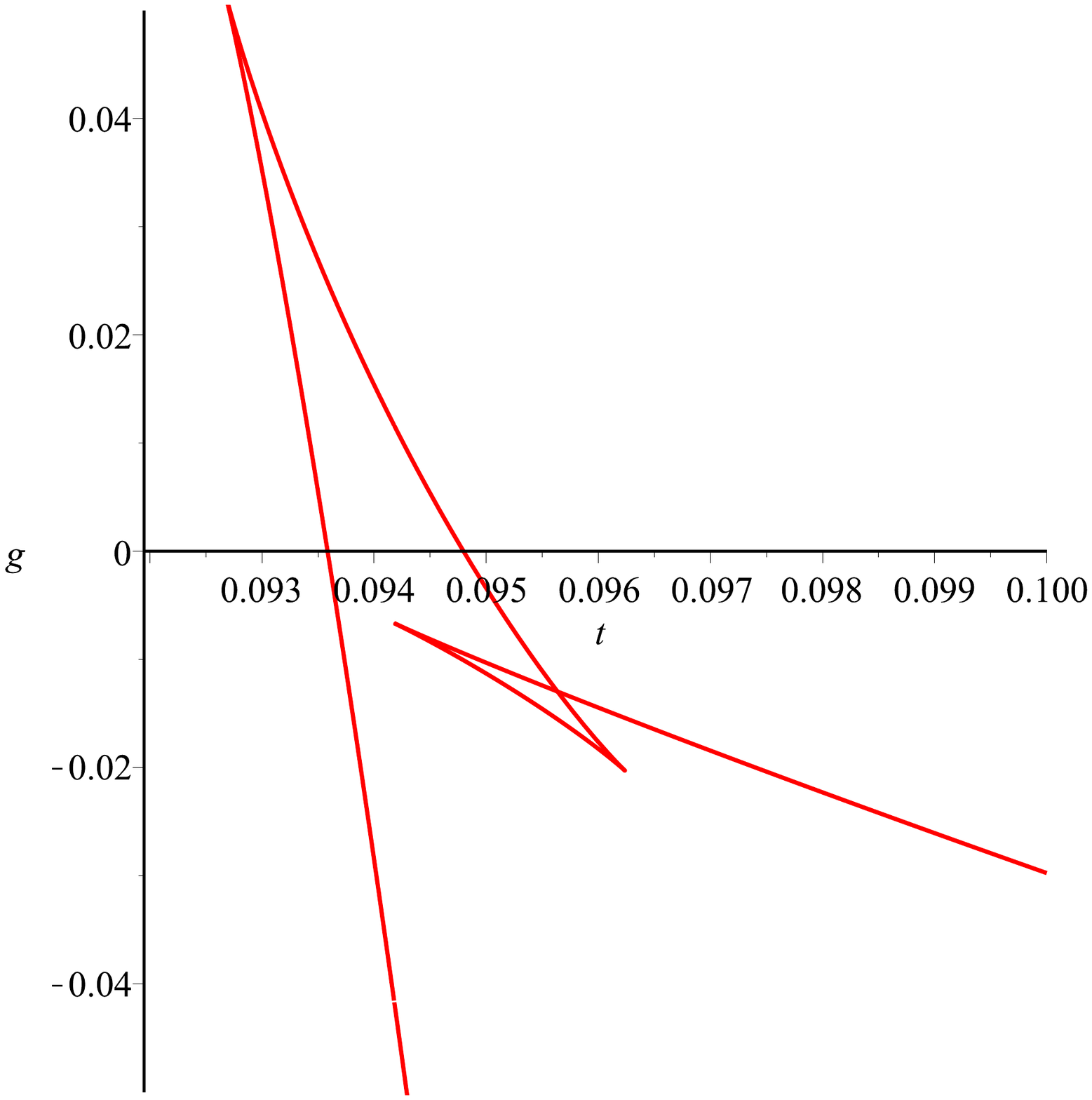}

\caption{{\bf Reentrant phase transition: $k=1$ case}. {\it Upper left}: $p-v$ plot for $\alpha =4,$ $ \tq=0,$ $ k=1$. The two black curves correspond to the two possible critical points; only the upper black curve corresponds to a physical critical point.  The blue curve is for $t=1.05t_c$, the red curve is for $t=0.96 t_c$, and the green curve is for $t=0.83 t_c$. {\it Upper right}: $g-t$ projection for $p=0.95p_c$ displaying classical swallowtail behaviour on the minimal branch, corresponding to a first order phase transition.  {\it Lower left}: $g-t$ projection for $p= 0.92 p_c$ showing the swallowtail no longer present on the minimal branch.  At $t_0$ minimization of the Gibbs free energy requires a discontinuous jump, indicating a zeroth order phase transition.  At $t_1$ the swallowtail intersects the minimal branch and here is a first order phase transition.  These two phase transitions combine to yield a reentrant phase transition. {\it Lower right}: $g-t$ projection for $p=0.8 p_c$.  The swallowtail is on the non-minimal branch and there is no intersection with the minimal branch.  There is no criticality in this region. 
}
\label{gt_a4_k1}
\end{figure*}
\end{center}

\begin{center}
\begin{figure*}[htbp]
  \includegraphics[width=.45\linewidth]{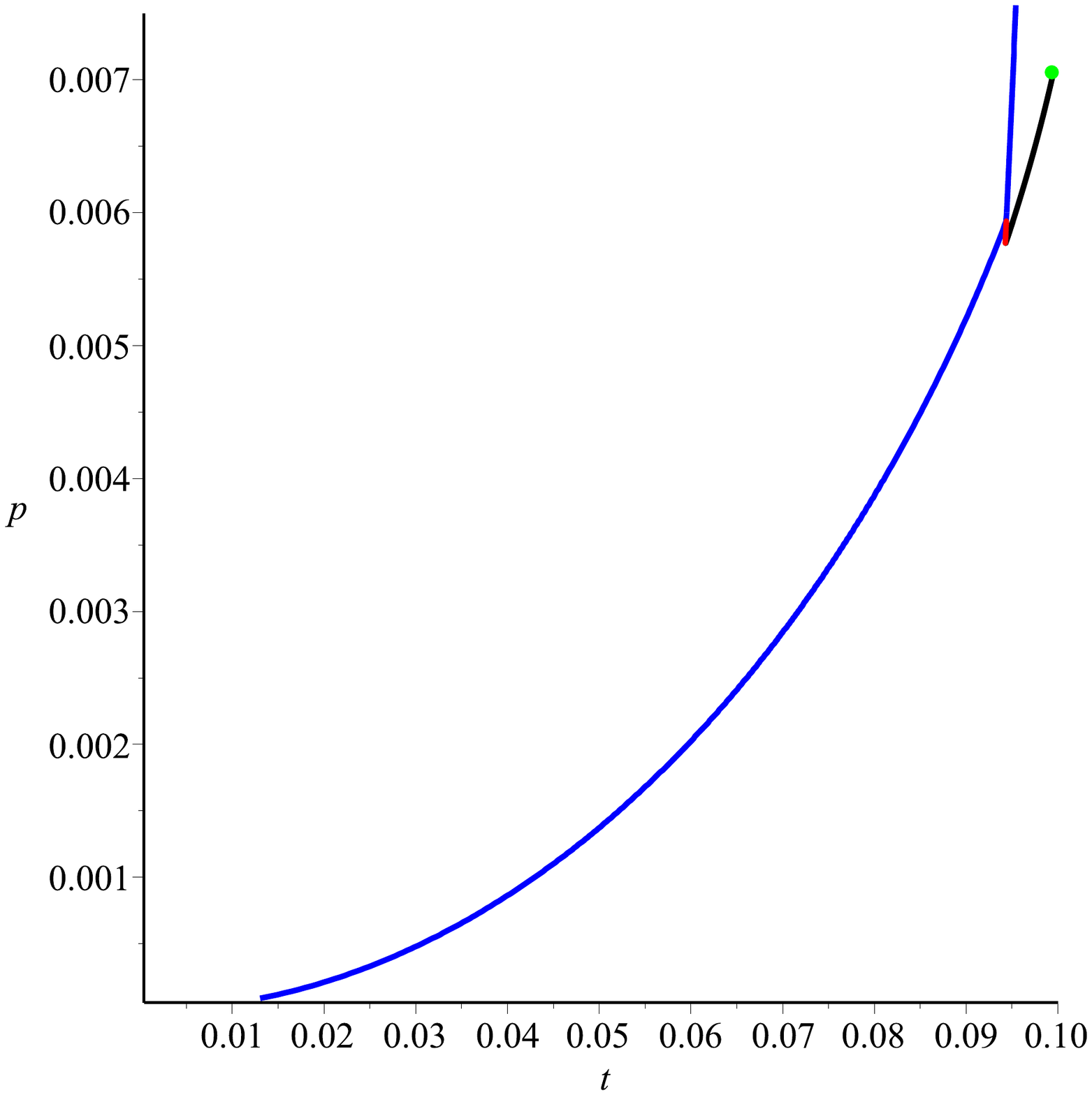}
\quad
  \includegraphics[width=.45\linewidth]{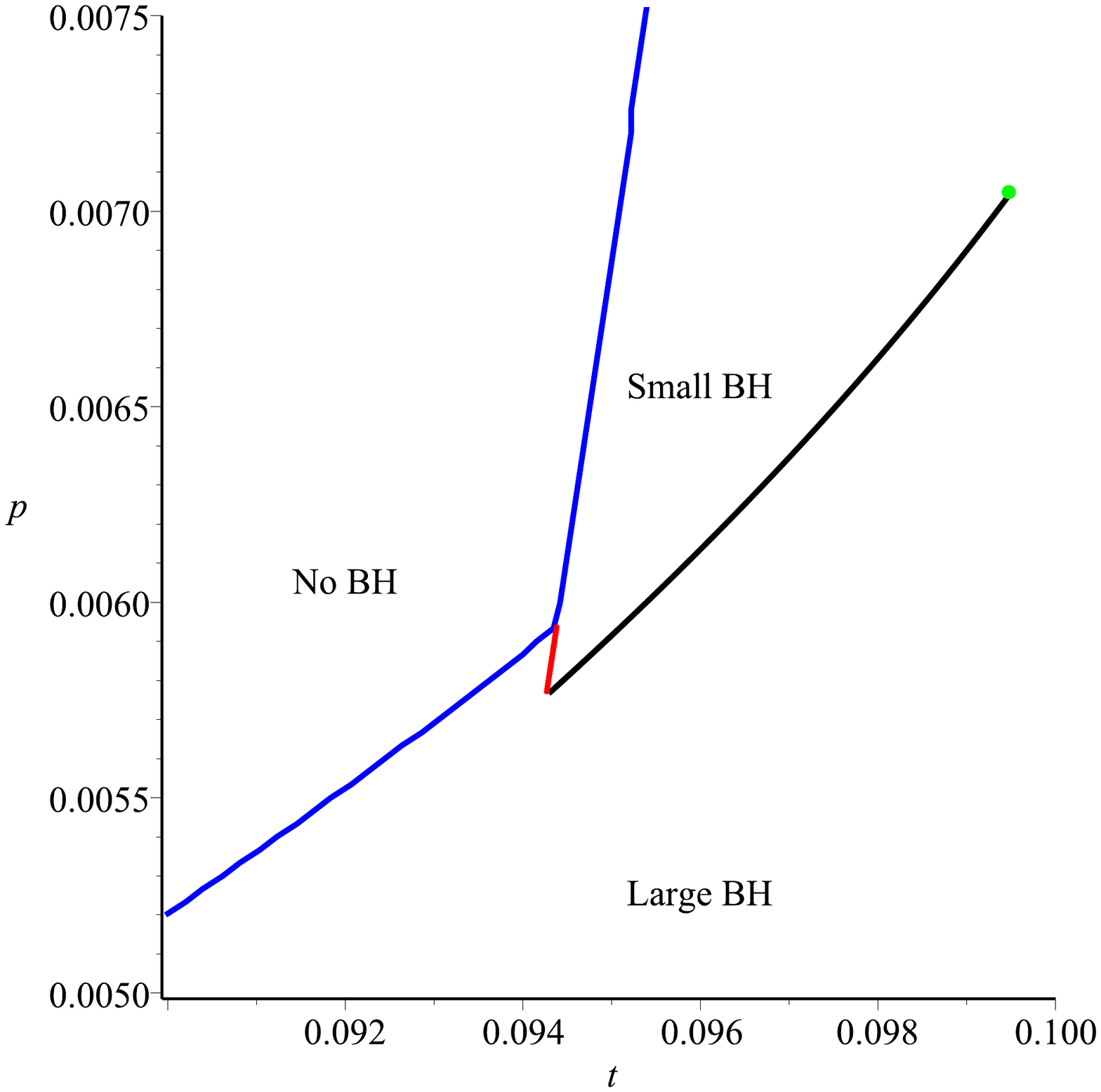}

\caption{{\bf Reentrant phase transition: $p-t$ plot for $\alpha = 4$, $k=1$}. The figure on the right zooms in on the region of interest.  
The coexistence line (black) starts at the virtual triple point and terminates at the physical critical point (green point).  A zeroth-order phase transition (red) begins at the virtual triple point and continues until it meets with the curve marking the boundary for the `no black hole region' (blue).  We see that in this case there is a large--small--large reentrant phase transition.}
\label{pt_a4_k1}
\end{figure*}
\end{center}

More interesting is the situation that occurs when $\alpha \in (\sqrt{225 + 120\sqrt{15}}/9, \sqrt{282 +102 \sqrt{17}}/6)$.  The behaviour of the Gibbs free energy and a $p-v$ plot is shown in Figure \ref{gt_a4_k1}, while Figure~\ref{pt_a4_k1} shows a representative $p-t$ plot.  In this case there are two candidates for critical points.  We find that there is a single physical critical point (characterized by the standard critical exponents), and the Gibbs free energy also shows a cusp.   We find that for pressures and temperatures lower than and close to the physical critical point the Gibbs free energy displays classical ``swallowtail" behaviour and the $p-v$ plot shows typical van der Waals behaviour.  

The swallowtail is only present on the minimal branch of the Gibbs free energy for a small range of pressure and temperature, after which it is present on the upper branch (see Figure~\ref{gt_a4_k1}). After swapping to the upper branch, the {tip of the} swallowtail intersects the minimal branch, resulting in both zeroth order and first order phase transitions. This combination of zeroth and first order phase transitions leads to a reentrant phase transition \cite{Altamirano2013}, so named
 because after multiple (in this case two) phase transitions, the system returns to the same macroscopic type of state upon
 a monotonic change in the temperature \cite{Hudson:1904}.

To see this, first consider the lower left plot in Figure~\ref{gt_a4_k1}. For a temperature $t < t_0$ the Gibbs energy is minimized for large black holes.  Increasing the temperature, the system undergoes a zeroth order phase transition to a small black hole at $t=t_0$ as there must be a discontinuous jump to minimize the Gibbs energy.   Further increasing the temperature, a first order phase transition occurs at $t=t_1$, and the system is again a large black hole.  In Figure~\ref{pt_a4_k1} we see the same effect -- for pressures in the range $p \in\,  \sim\!( 0.0057828, 0.0059372)$, a large--small--large reentrant phase transition is possible.  The phase behaviour terminates at the point where the zeroth and first order phase transitions coincide in what is known as a ``virtual triple point''.  In all cases the critical behaviour is consistent with the maximal pressure constraint. 

In the region $\alpha > \sqrt{282 +102 \sqrt{17}}/6$ the single candidate for a critical point is  similar to the one just described, and the plots presented in the previous discussion are qualitatively identical to what is observed  for these values of $\alpha$.

\begin{center}
\begin{figure}[htp]
  \includegraphics[width=.9\linewidth]{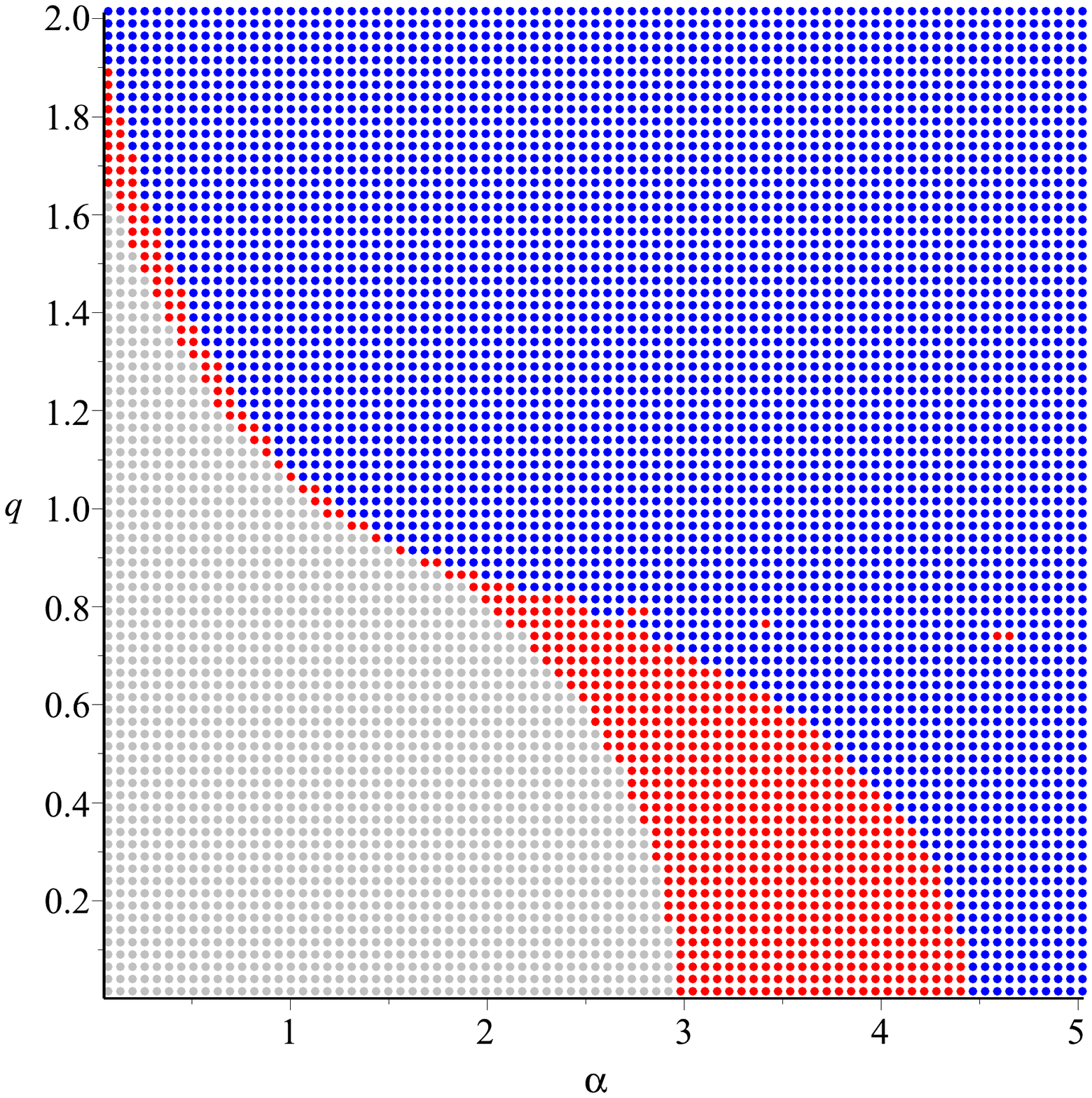}
\\
  \includegraphics[width=.9\linewidth]{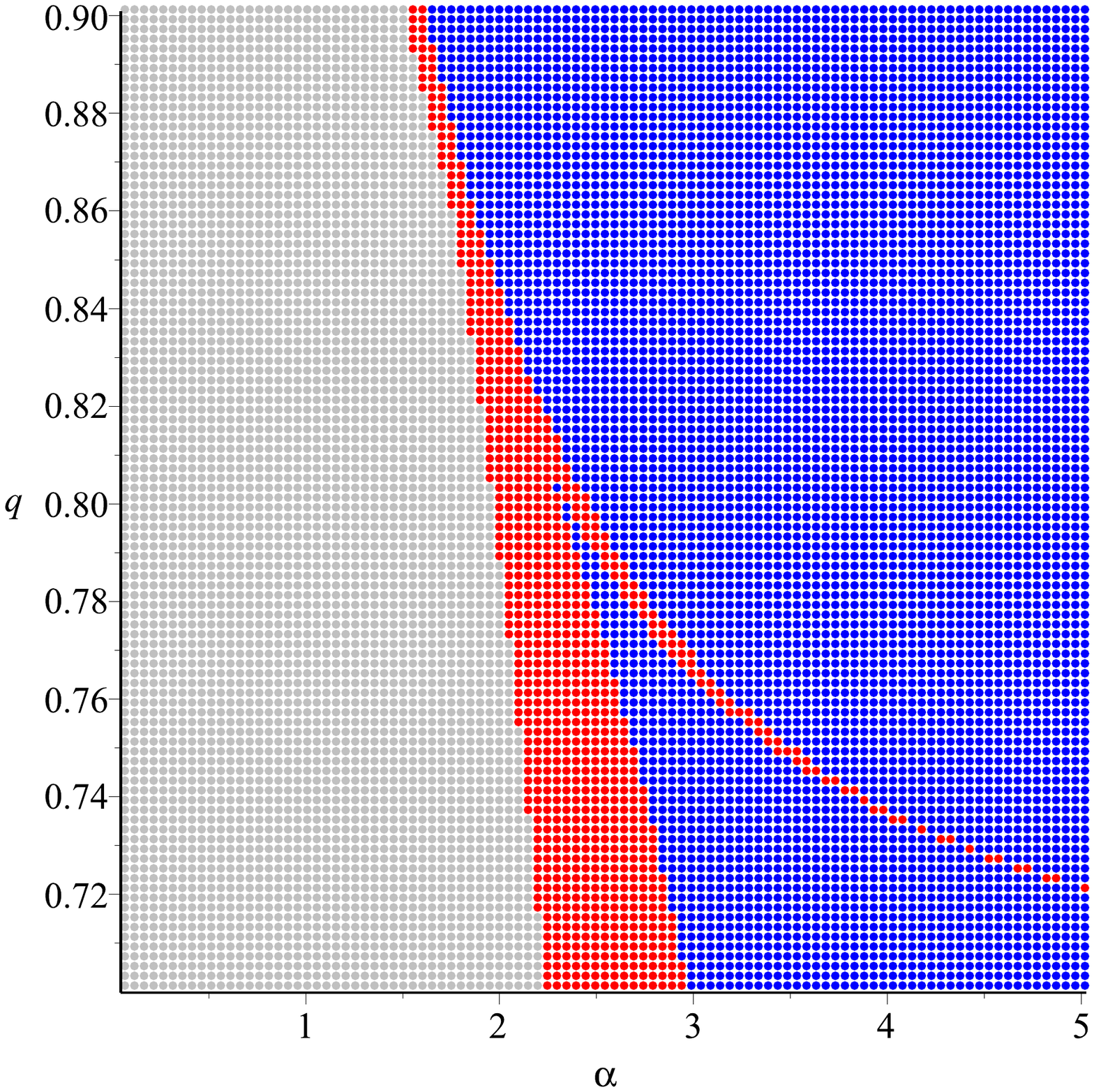}
\caption{{\bf Critical points in $(\alpha, \tq)$ parameter space for $D=5$ and $k=1$}. {\it Top}: Possible critical points with $(p_c, v_c, t_c)$ satisfying pressure and entropy constraints.  Gray dots indicate zero critical points, blue dots indicate a single critical point, and red dots indicate two critical points.  {\it Bottom}: A close up view of the top plot focusing on $0.7 \le \tq \le 0.9$ showing more detail. }
\label{qa_scan_5d_k1}
\end{figure}
\end{center}

\begin{center}
\begin{figure}[htp]
\includegraphics[width=.8\linewidth]{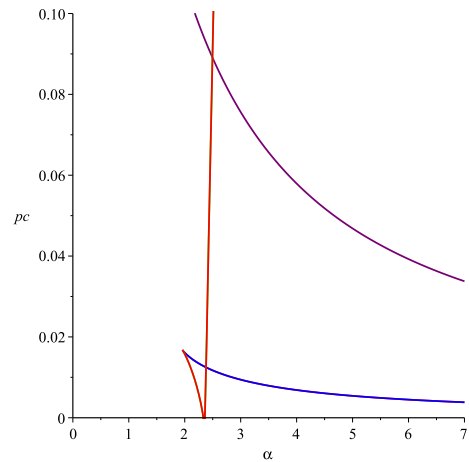}
\caption{A plot of $p_c$ vs. $\alpha$ for $D=5$, $k=1$ and $\tq = 0.795$.  We see that at the intersection of
the red and blue curves ($\alpha \approx 2.3834328996$) the two critical pressures coincide.  The purple curve is the maximum pressure constraint. }
\label{pc_vs_alpha_k1_5d}
\end{figure}
\end{center}

\begin{figure*}[htp]
\centering
\includegraphics[width=0.3 \linewidth]{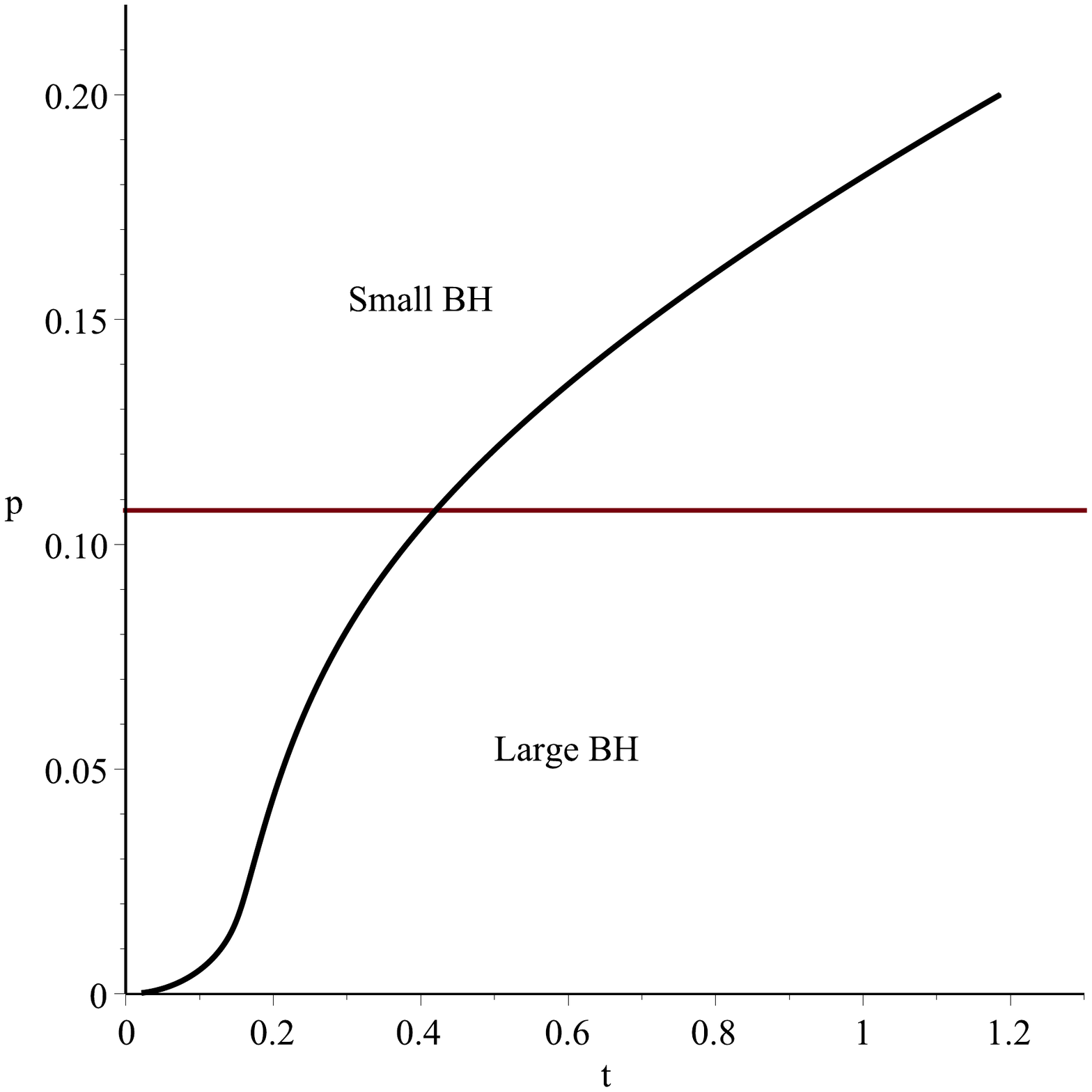}
 \includegraphics[width=0.3\linewidth]{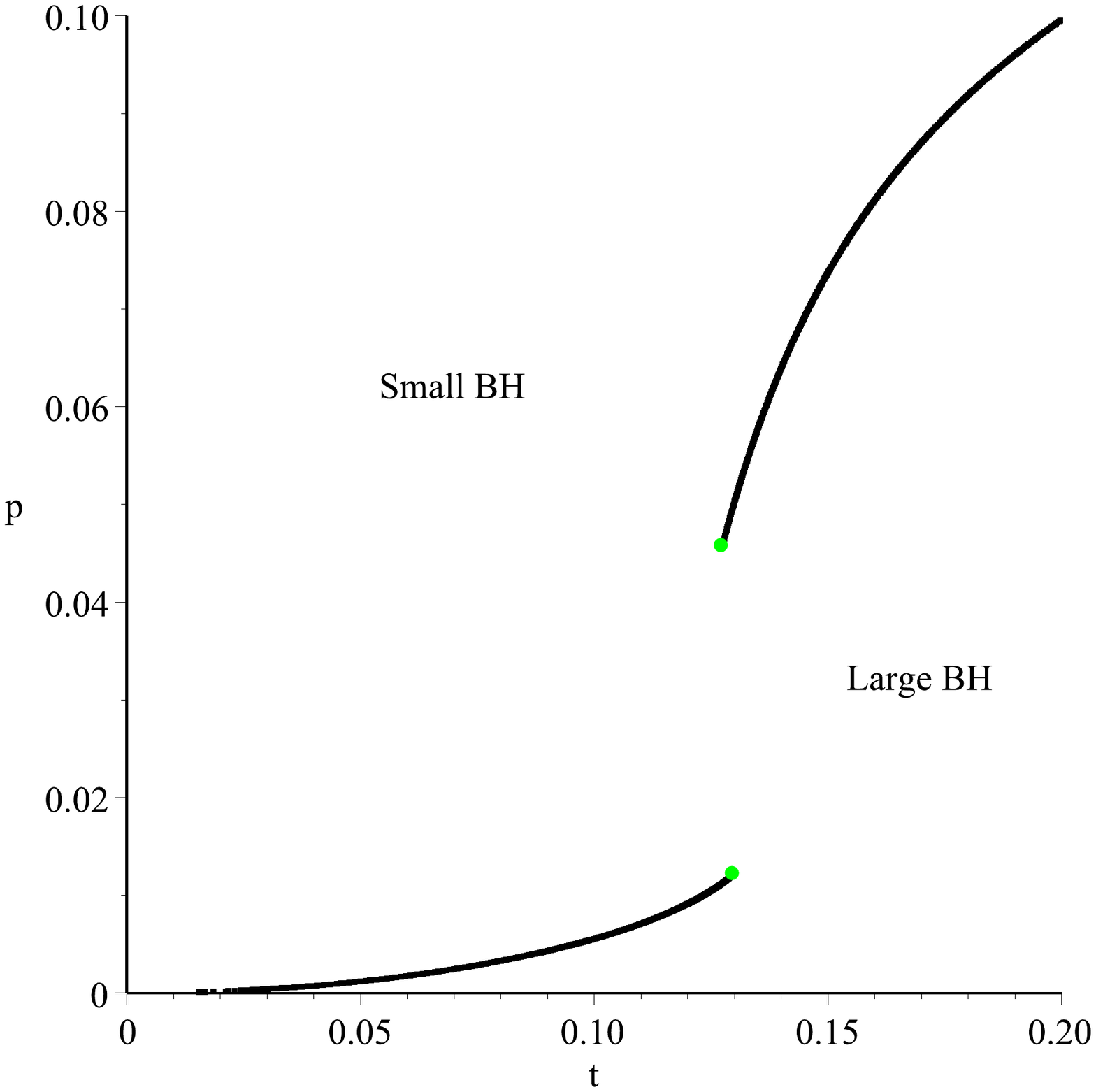}
  \includegraphics[width=0.3\linewidth]{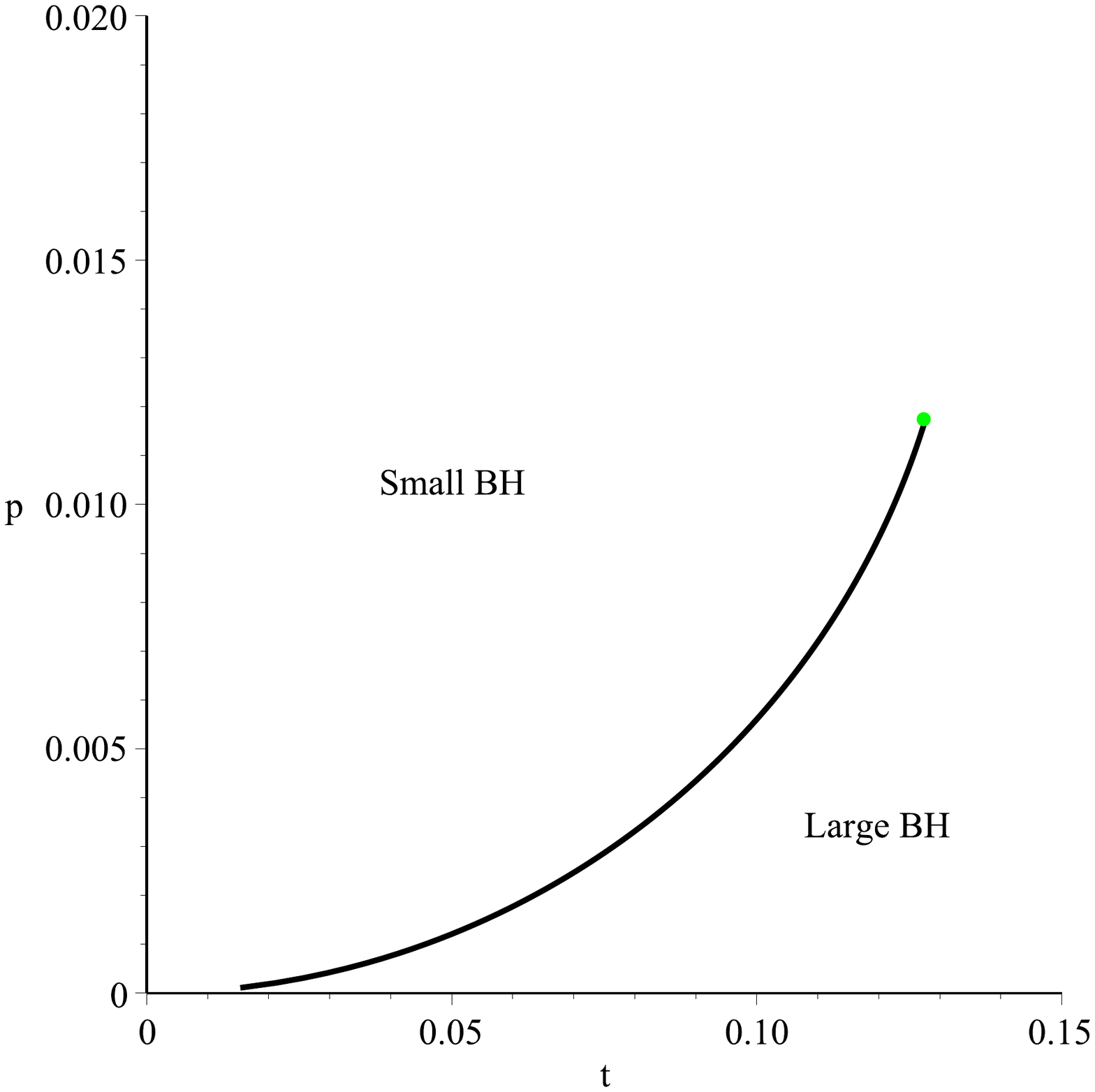}

\caption{{\bf Charged case: $p-t$ plots for $\tq = 0.795$, $k=1$, $D=5$}.  {\it Left}: $p-t$ plot for $\alpha = 2$ showing the absence of physical critical points and a coexistence line indicating a first order phase transition between small and large black holes.    {\it Center}: $p-t$ plot for $\alpha = 2.43$ here we see two physical critical points (green dots).  The lower coexistence line displays standard van der Waals behavior: beginning at $(0,0)$ and terminating at a critical point.  The upper coexistence line displays `reverse van der Waals' behaviour: beginning at a critical point and continuing on indefinitely.  {\it Right}: $p-t$ plot for $\alpha=2.5$ showing only a single physical critical point and van der Waals-type behaviour.  }
\label{pt-plots_q_k1_5d}
\end{figure*}

\begin{center}
\begin{figure*}[htp]
  \includegraphics[width=.35\linewidth]{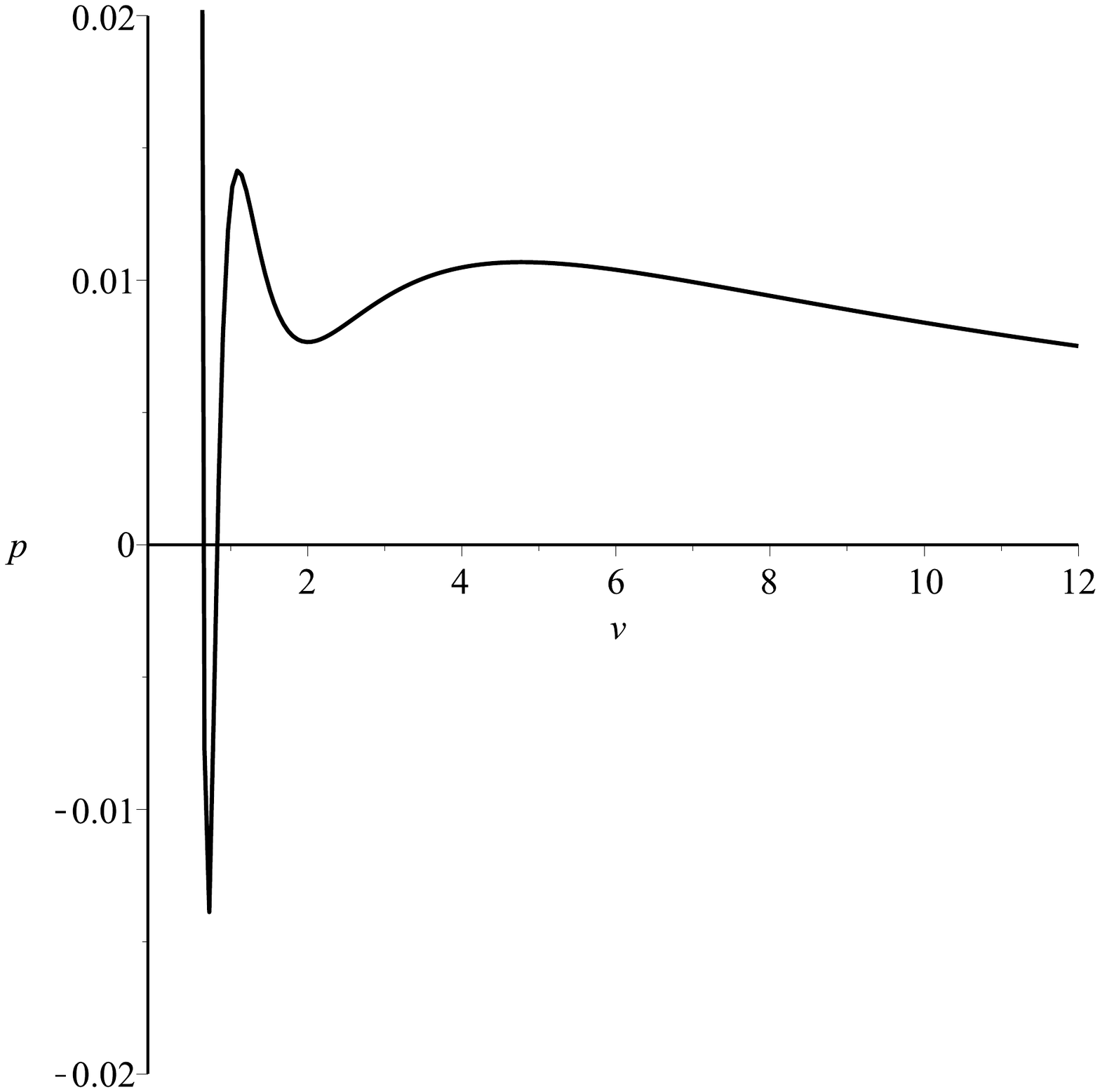}
  \includegraphics[width=.35\linewidth]{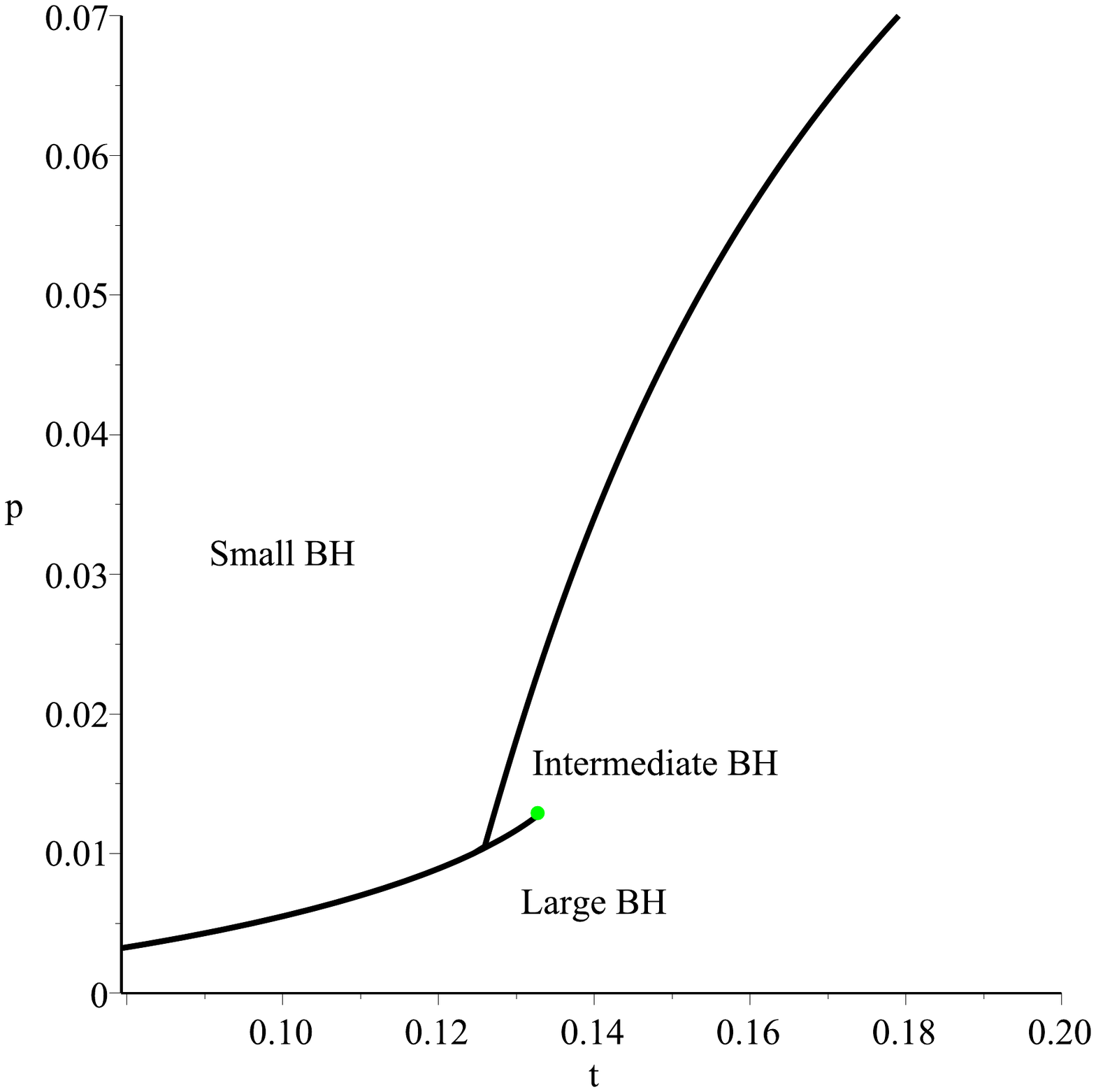}
  \includegraphics[width=.35\linewidth]{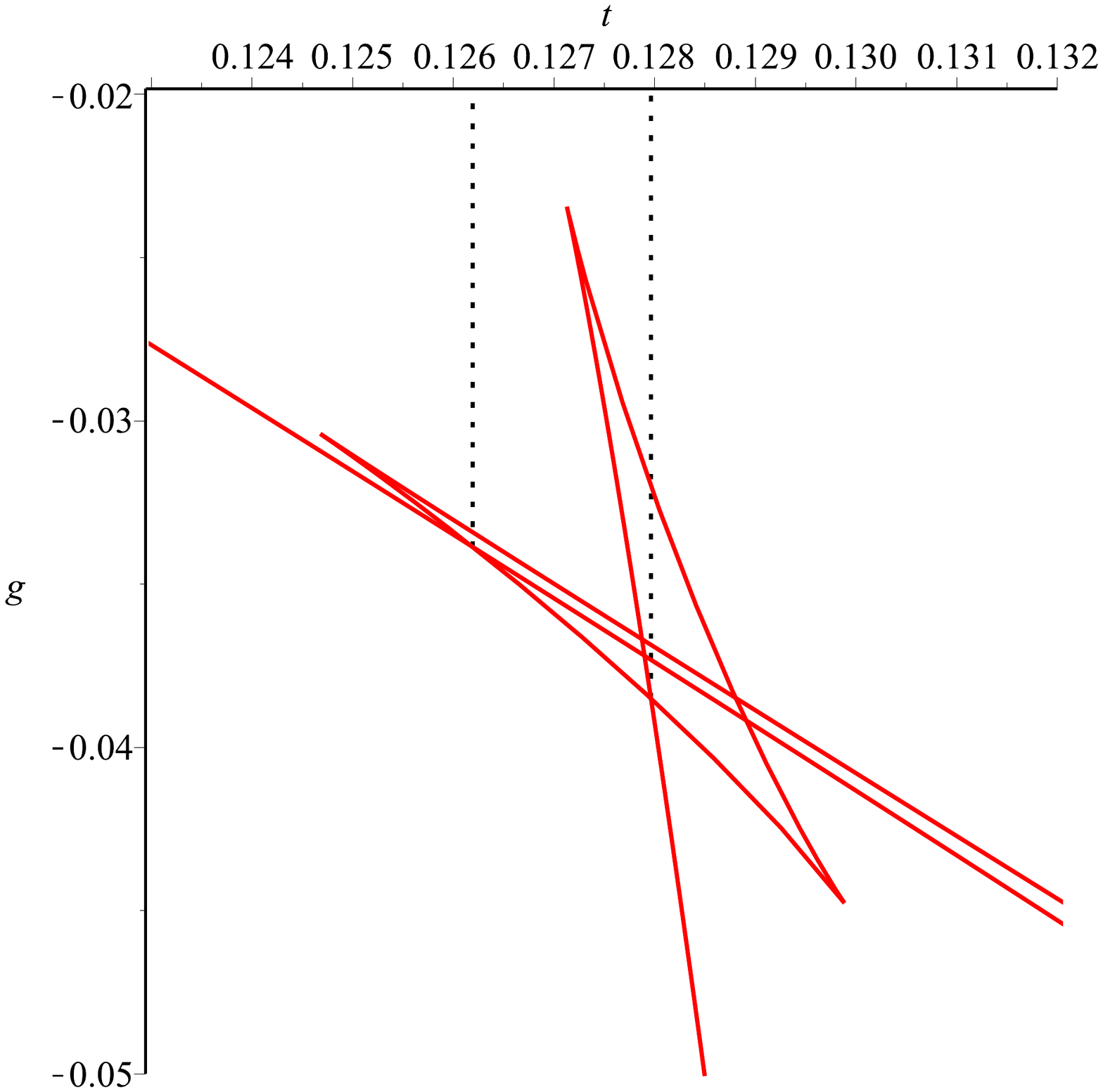}
  \includegraphics[width=.35\linewidth]{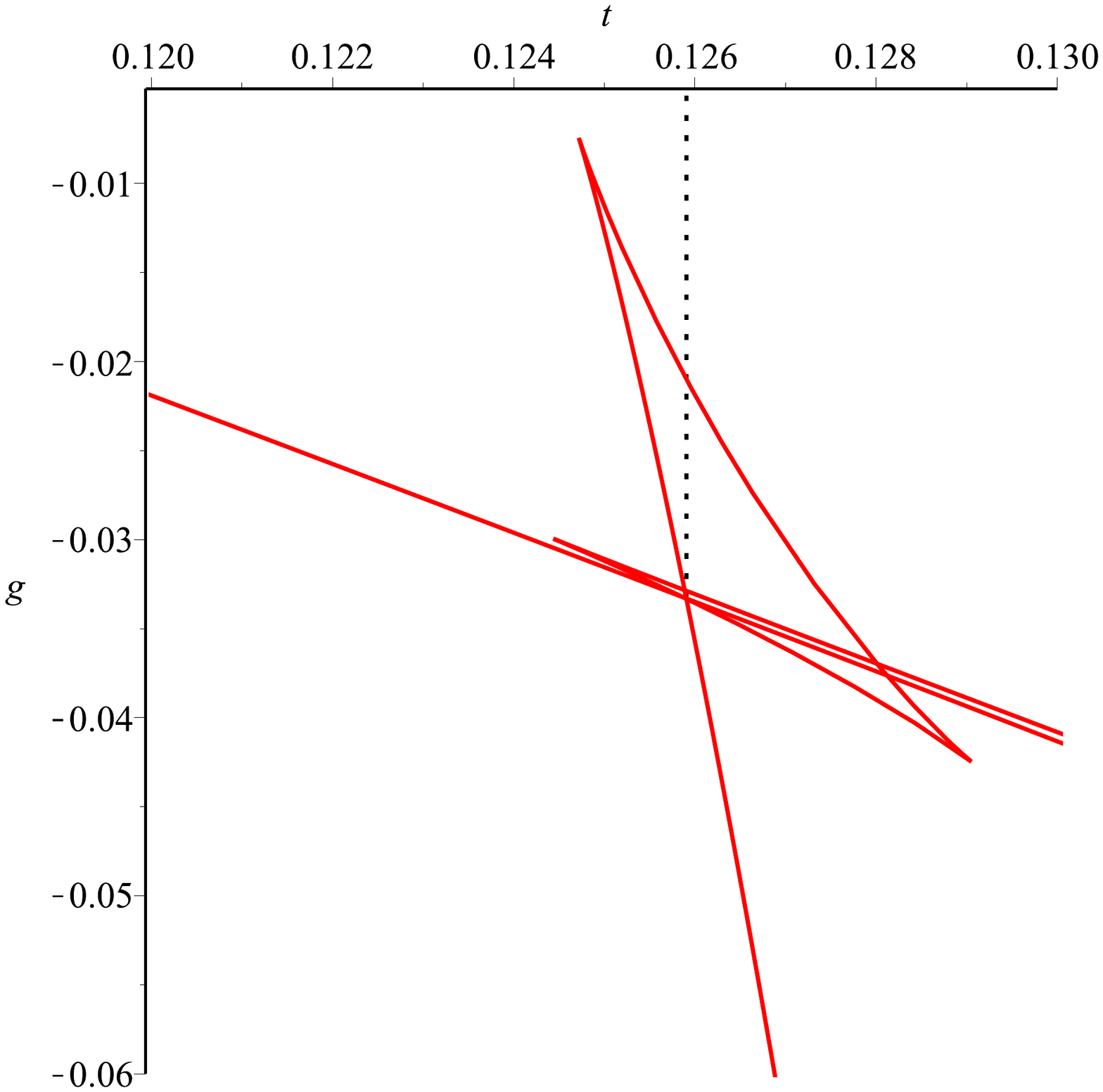}

\caption{{\bf Triple point:} $\tq = 0.795$, $\alpha = 2.33$, $k=1$, $D=5$. {\it Upper left}: $p-v$ plot at the triple point.   {\it Upper right}: $p-t$ plot for the triple point.  Note the presence of two first order phase transition coexistence lines distinguishing small, intermediate and large black holes.  At the triple point the coexistence lines meet.  The lower coexistence line terminates at a critical point (green dot), while the upper coexistence line continues indefinitely, eventually becoming unphysical when it crosses the maximal pressure constraint (greater than the values of $p$ shown on the plot).  {\it Lower left}: $g-t$ projection for $p > p_{3c}$ showing the presence of two first order phase transitions, indicated by the black dotted line.  {\it Lower right}: $g-t$ projection for $p = p_{3c}$.  At the triple point the two first order phase transitions merge together. }
\label{k1_5d_triple_point}
\end{figure*}
\end{center}

We now turn our attention to the charged case.  To begin with, we study the $(\tq, \alpha)$ parameter space to determine the number of critical points for  particular combinations of $\tq$ and $\alpha$.  The results of this investigation are presented in Figure~\ref{qa_scan_5d_k1}.  There are no physical critical points for $\alpha < 0$.   For $\alpha>0$, depending on the $\tq$, $\alpha$ combination there may be as many as two critical points. 

Figure~\ref{pc_vs_alpha_k1_5d}  elucidates the thermodynamics in the case of two critical points.  
We present a representative analysis for the case $\tq = 0.795$.  
Here we observe an interesting feature for the critical pressure, namely that for  $\alpha~\approx~1.958731054$ and $\alpha~\approx~2.3834328996$ two critical pressures coincide, as can be seen in Figure~\ref{pc_vs_alpha_k1_5d}.  For $\alpha~\approx~1.958731054$ the two critical points coincide, while for $\alpha~\approx~2.3834328996$ only the critical pressures coincide, and the critical temperatures and volumes remain distinct.  In each case, the critical point is governed by the standard mean field theory critical exponents.  Hence these are not isolated critical points like that found in \cite{Frassino2014,Dolan:2014vba}.

For $\alpha \in \,\sim\!\! (1.95, 2.23)$ neither of the two possible critical points are physical as they occur on a non-minimal branch of the Gibbs free energy.  In this region we see an infinite coexistence line separating small and large black holes, as shown in Figure~\ref{pt-plots_q_k1_5d}. For $\alpha  \gtrsim 2.23$ we see the appearance of two separate physical critical points.  A coexistence line beginning at the origin terminates at one of these critical points, while a coexistence line emanates from the second critical point, extending beyond physical pressures.  As $\alpha$ is increased beyond $\sim 2.23$ these physical critical points occur for more widely separated pressures and eventually one of the critical points becomes unphysical -- exceeding the maximal pressure constraint.   

For a small range of $\alpha  \in \, \sim\!\! (2.23, 2.38) $  we find a triple point.  This feature is presented for  $\alpha = 2.33$ in Figure~\ref{k1_5d_triple_point}.  Referring to the $p-v$ plot, we see the `double van der Waals' oscillation that is characteristic of a triple point.   In the $p-t$ plane we see two coexistence lines for first order phase transitions that come together at the triple point.  Near the triple point we see the presence of a small/intermediate/large black hole phase transition that is reminiscent of the solid/liquid/gas phase transition.  Considering the Gibbs free energy, we see in Figure~\ref{k1_5d_triple_point}c that for pressures slightly above the triple point there are two first order phase transitions.  As the pressure is decreased, these first order phase transitions come closer together until they finally merge at the triple point, as is shown in Figure~\ref{k1_5d_triple_point}d.

\begin{center}
\begin{figure}[htp]
\includegraphics[width=.8\linewidth]{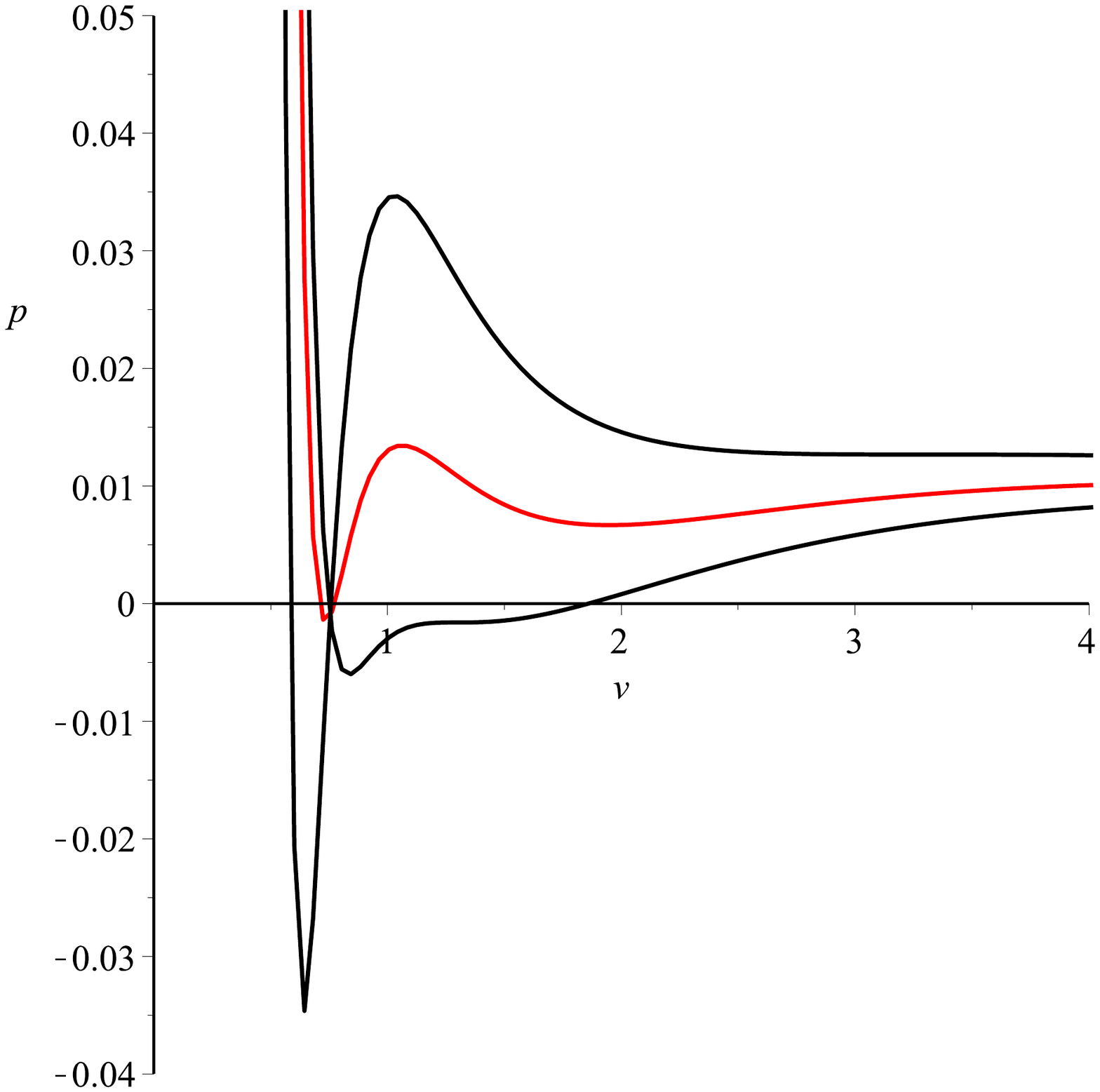}
\caption{A $p-v$ plot for $\tq = 0.795$, $k=1$, $D=5$, $\alpha= 2.4$ showing multiple isotherms intersecting at a single point, indicating the presence of a thermodynamic singularity. }
\label{5d_k1_q_td_sing}
\end{figure}
\end{center}

As discussed earlier, the $k=1$ thermodynamic singularity can occur for physical pressures when charge is included.  One may wonder if the thermodynamic singularity ever coincides with a critical point.  Simple calculations show that provided 
\be 
\alpha = \frac{1}{2}\sqrt{\frac{36 - 2 \pi^2 \tq^4 - 12 \pi \tq^2 + 2\pi \tq^3 \sqrt{\pi(\pi \tq^2 + 12)}}{2\pi \tq^2 - 3}}
\ee
then the thermodynamic singularity occurs at a potential critical point; an analysis of its features indicates
that it is an isolated critical point.  However we find that the maximal pressure constraint is always violated for the value of $\alpha$ at which the thermodynamic singularity coincides with the critical point. Hence this   this critical point is always unphysical.   

When the thermodynamic singularity is physical it appears as a point at which all of the isotherms intersect at a particular volume.  This feature is highlighted in Figure~\ref{5d_k1_q_td_sing} for $\alpha = 2.4$ and $\tq = 0.795$.  Evaluating eqs.~\eqref{tdsingpoint} for the these parameter values we find that,
\ba 
p_s \approx 0.02835&&105681, \quad t_s \approx 0.1267089635, 
\nn\\
 v_s &&\approx 0.7481508651.
\ea
Expanding the Gibbs free energy and temperature about the thermodynamic singular point we find that,
\ba 
g(v_s + dv, p_s + dp) &=& A \frac{dp}{dv} + B + \mathcal{O}(dp, dv)
\nn\\
t(v_s + dv, p_s + dp) &=& C \frac{dp}{dv} +t_s + \mathcal{O}(dp, dv)
\ea
where $A, B$ and $C$ are numerical constants.  We observe that the Gibbs free energy expressed as a function of $v$ and $p$ does indeed diverge at the thermodynamic singularity.  However, as noted in~\cite{Frassino2014}, the Gibbs free energy is more naturally expressed as a function of $t$ and $p$.  In terms of these natural variables, the expansion of the Gibbs free energy near the thermodynamic singularity is,
\be 
g(t_s + dt, p_s + dp) = \frac{A}{C} dt + B + \mathcal{O}(dp, dv)
\ee
which is finite and smooth.  We therefore conclude that the thermodynamics are well-behaved at the thermodynamic singular point.  We note that when the thermodynamic singularity does not occur at a potential critical point (which is the case for the physical critical points considered here), the system is governed by the standard mean field theory critical exponents given by eq. \eqref{critExps}.

\subsubsection{Hyperbolic}

In the case $k=-1$ we find that, regardless of the value of $\tq$, for both positive and negative $\alpha$ there is only one possible critical point (positive $v_c, t_c$ and $p_c$).  However, in all cases this critical point is unphysical, since we find $p_c$ exceeds the maximum pressure constraints.  This being the case, we still observe interesting thermodynamic phenomena.  
\begin{center}
\begin{figure}[htp]
\includegraphics[width=.8\linewidth]{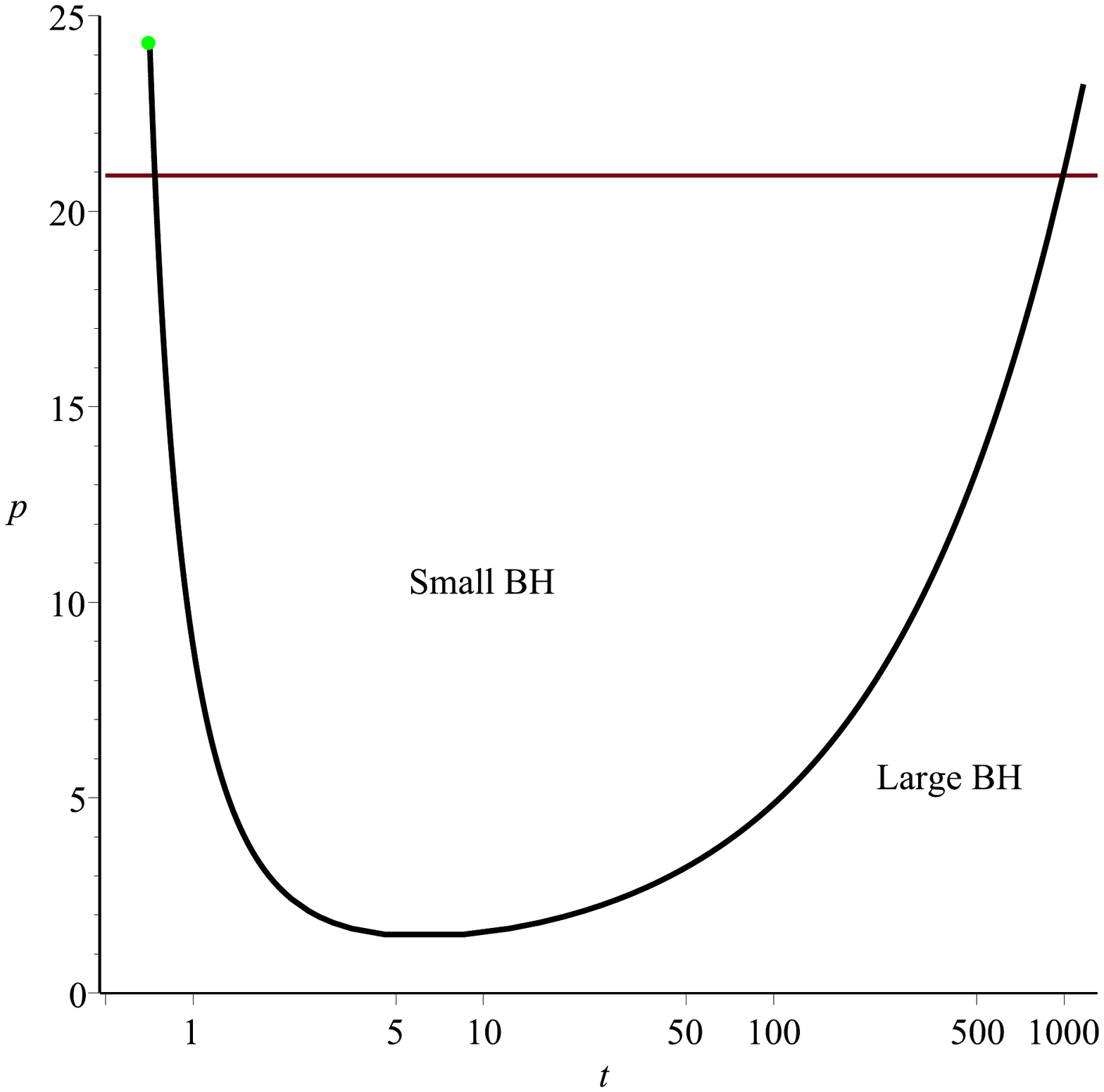}
\caption{A representative $p-t$ plot using $\alpha =-5$, $\tq=0$, $k=-1$ showing the presence of a large/small/large reentrant phase transition.  The horizontal red line corresponds to the maximal pressure constraint.  The coexistence line originates at an unphysical critical point (green dot), but the coexistence line itself extends into the physical regime for a large range of temperatures and pressures.  The temperature axis has been represented logarithmically to highlight the presence of the reentrant phase transition.  }
\label{km1_pt_am5}
\end{figure}
\end{center}

For $\alpha < 0$,  positivity of entropy is automatically satisfied.  Here we observe the presence of a large/small/large reentrant phase transition, as can be seen in Figure~\ref{km1_pt_am5} which shows a representative $p-t$ plot for $\alpha=-5$ and $\tq=0$.  We find that the actual critical point is unphysical: it occurs at a pressure exceeding the maximum pressure constraint.  However, the coexistence line that begins at this critical point dips below the maximum pressure constraint, allowing for physical first order phase transitions.  Since the coexistence line forms a ``U" shape, it is possible for two successive first order phase transitions to occur, resulting in a reentrant phase transition.  Note that a qualitatively similar $p-t$ plot is found for all $\alpha < 0$ and for $\tq > 0$.  

\begin{center}
\begin{figure*}[htp]
 \includegraphics[width=.32\linewidth]{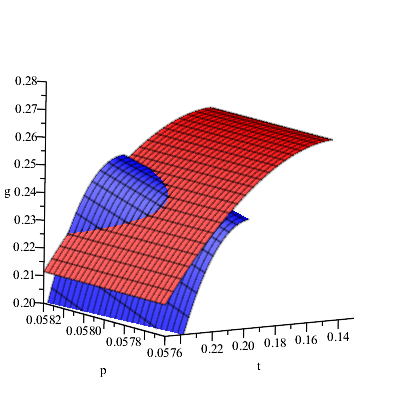}
  \includegraphics[width=.3\linewidth]{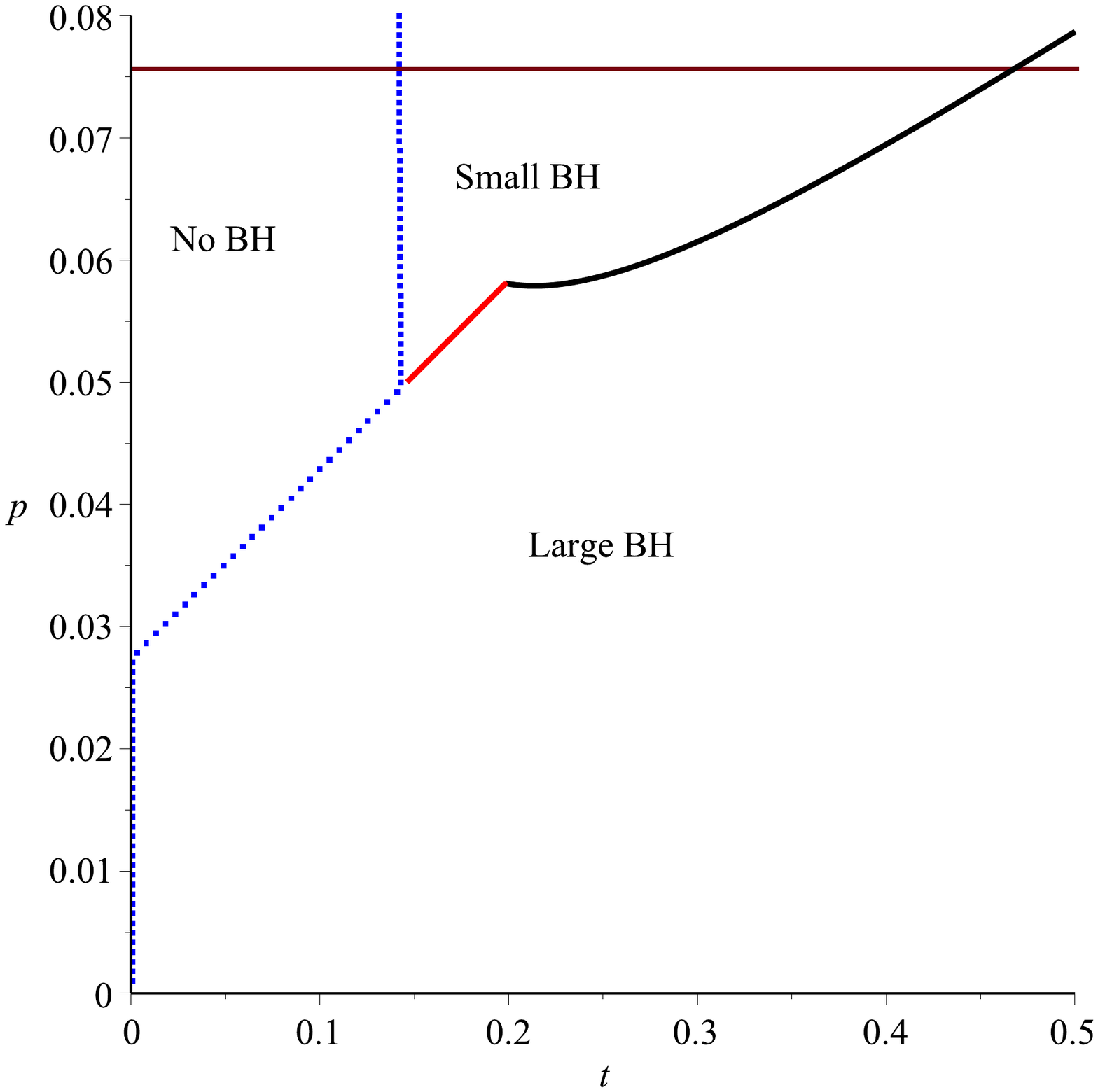}
  \includegraphics[width=.3\linewidth]{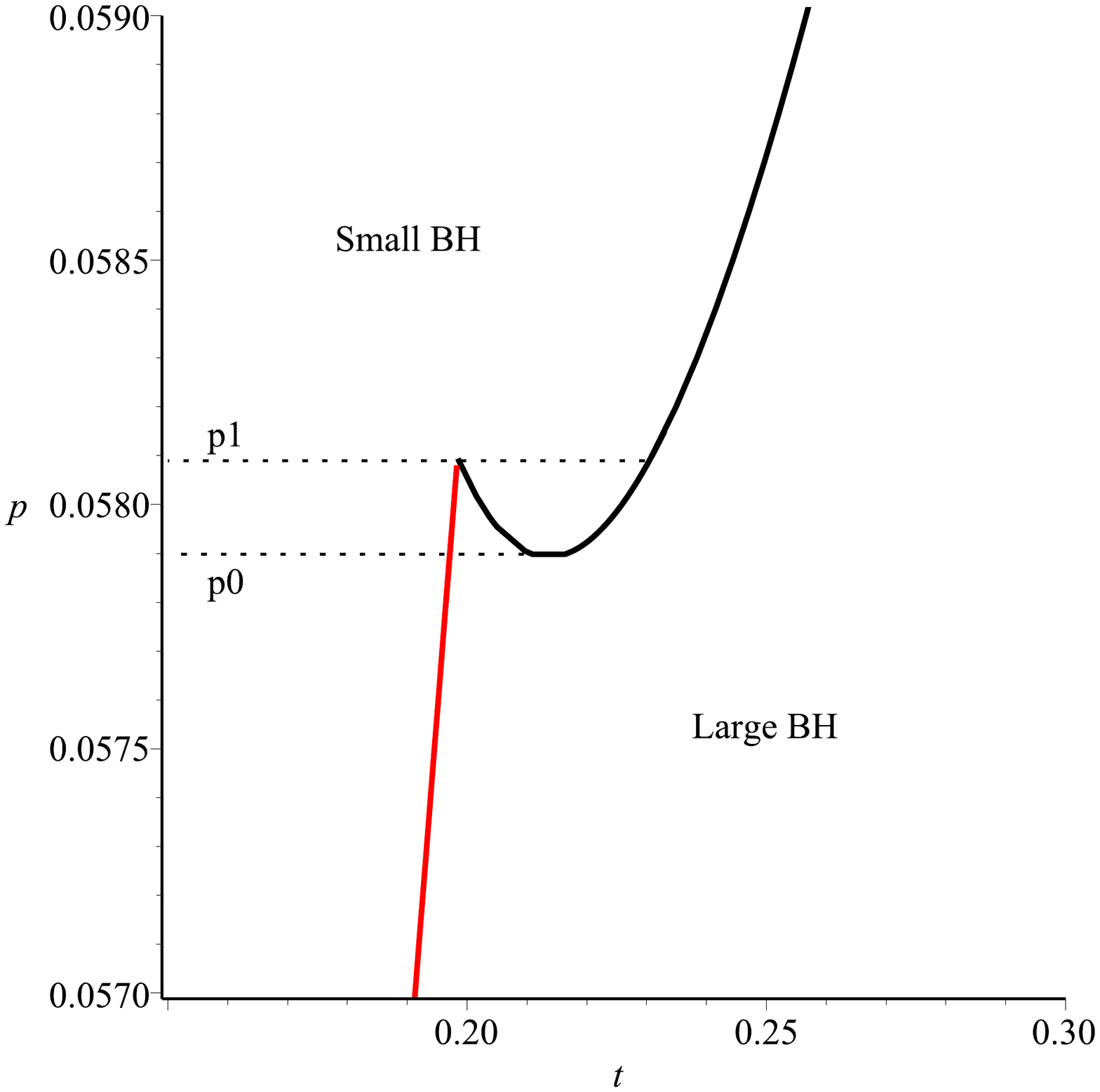}

\caption{{\bf Multiple reentrant phase transition:} $\tq = 0$, $\alpha = 3$, $k=-1$, $D=5$. {\it Left}: A plot of the Gibbs free energy in the parameter range where multiple reentrant phase transitions are present.  {\it Center}: $p-t$ phase plot.  The solid black line represents the coexistence line associated with a first order phase transition, the solid red line represents a zeroth-order phase transition.  To the left of the dotted blue line there can exist no physical black holes (those with positive entropy).  The maximal pressure constraint is shown as the dark red horizontal line.  {\it Right}: A zoomed in view of the center figure showing the presence of a small/large/small/large multiple reentrant phase transition for pressures between $p_0$ and $p_1$.   }
\label{5d_mrpt}
\end{figure*}
\end{center}

When considering $\alpha > 0$, the positivity of entropy condition is no longer trivial and must be enforced by ensuring that eq. \eqref{posEntNegMu} is satisfied for $k=-1$.  For $\alpha  < \, \sim\! 1.000$ we observe features qualitatively similar to the $\alpha < 0$ case, and Figure~\ref{km1_pt_am5} suffices to display the physics at play.  However, as $\alpha$ becomes larger (still regarding black holes with negative entropy as unphysical), discontinuities are introduced into the Gibbs free energy.  These discontinuities give rise to zeroth order phase transitions which lead to  multiple reentrant phase transitions.  This feature is illustrated in Figure~\ref{5d_mrpt} for $\tq=0$ and $\alpha = 3$. 

Referring to Figure~\ref{5d_mrpt}(a) we see how the structure of the Gibbs free energy gives rise to  a multiple reentrant phase transition.  Due to the positivity of entropy constraints, there are only two physical branches of the Gibbs free energy, and these terminate at finite temperature.  Thus up to a certain temperature, there are no physical black holes.  For smaller pressures, we observe a zeroth order phase transition as there is a discontinuous jump    
 required to minimize the Gibbs free energy as the temperature increases.  For larger pressures, the two branches of the Gibbs free energy intersect twice giving rise to two first order phase transitions.  At a larger pressure ($ p \approx 0.0581$) the zeroth order phase transition coincides with the first order phase transition that occurs at the lowest temperature.  Above this pressure we observe only a single first order phase transition.   
 
Now considering Figure~\ref{5d_mrpt}(c) we see that for a fixed pressure $p$ in the region $(p_0, p_1)$ a monotonic variation of the temperature will result in a zeroth order small/large phase transition followed by a first order large/small phase transition and finally a first order small/large phase transition.  Thus we have a two reentrant phase transitions (small/large/small and large/small/large).  An identical result was found in \cite{Frassino2014} for $7D$ $3^{rd}$-order Lovelock gravity.  This result is the first example of multiple reentrant phase transitions for black holes in fewer than $7$ dimensions.  

\begin{figure*}[htp]
  \includegraphics[width=.4\linewidth]{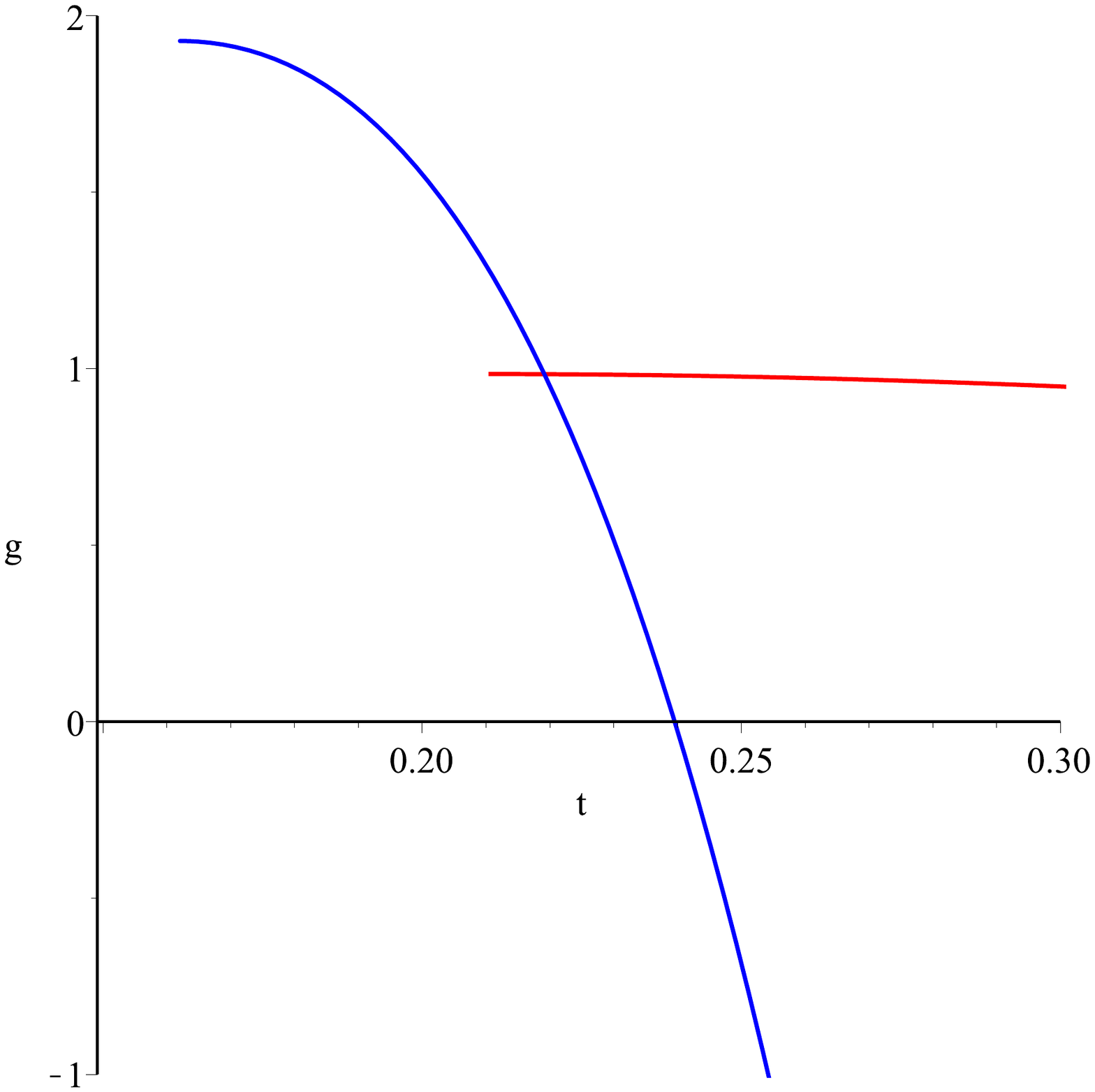}
  \includegraphics[width=.4\linewidth]{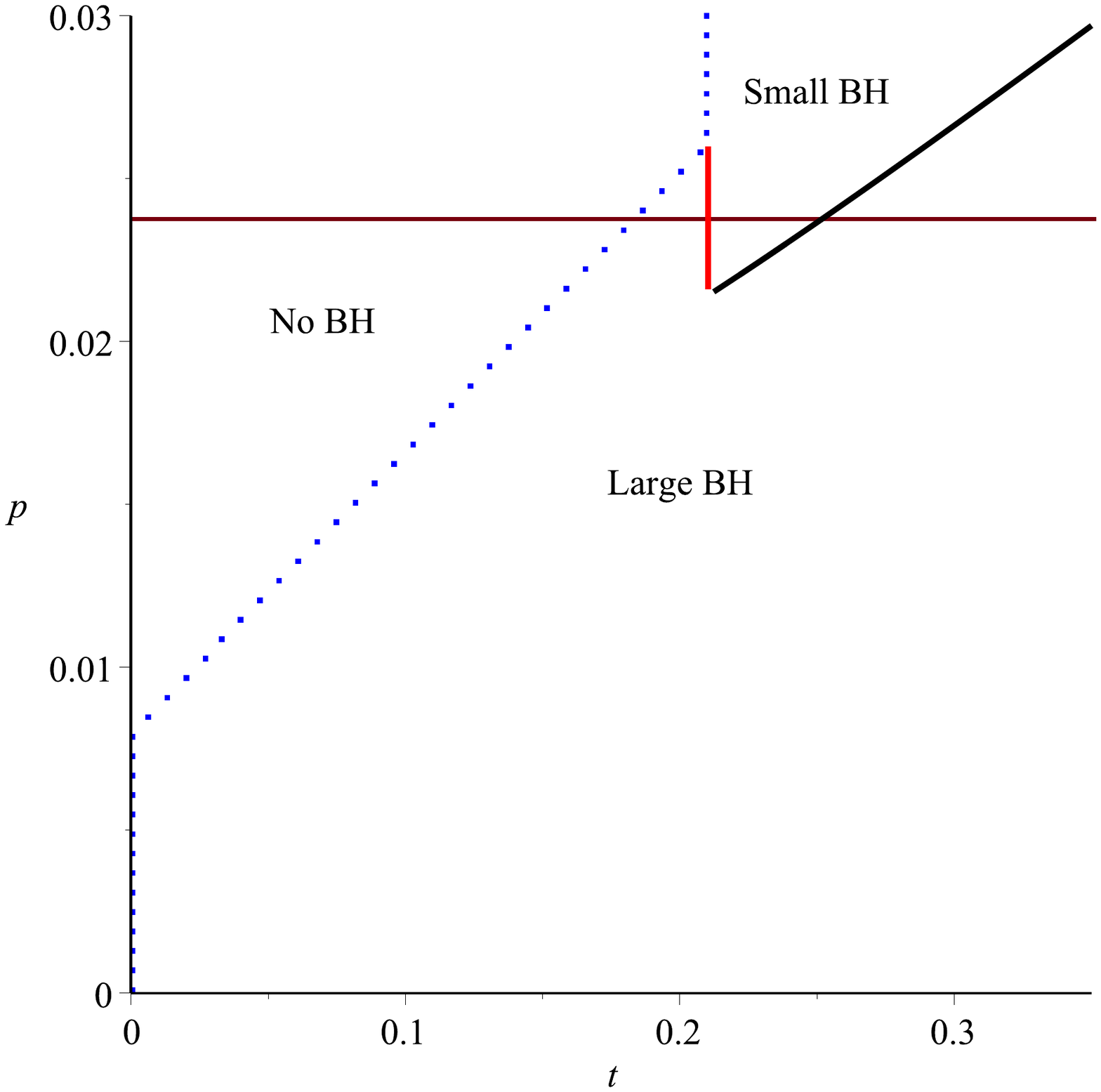}

\caption{{\bf Reentrant phase transition  $\alpha =10$, $\tq=0$, $k=-1$}: {\it Left}: A $g-t$ projection for $p=0.0219$.  Enforcing the positive entropy constraint leaves only two physical branches of the Gibbs free energy, both of which begin at non-zero temperature, indicated a `no black hole' region. {\it Right}: A representative $p-t$ plot for showing the presence of a large/small/large reentrant phase transition.  The horizontal dark red line corresponds to the maximal pressure constraint.  A first order phase transition (black line) and a zeroth order phase transition (red line) are present.  The dotted blue line marks the region where the black holes have negative entropy (and hence are regarded as unphysical).     }
\label{5d_pt_a10_q0}
\end{figure*}

By further increasing $\alpha$ we find that the range of pressure for which multiple reentrant phase transitions occurs  decreases until eventually at $\alpha \,\sim  4.916$ we cease to observe multiple reentrant phase transitions.  For $\alpha \gtrsim  4.916 $ we observe a single large/small/large reentrant phase transition, as shown in Figure~\ref{5d_pt_a10_q0}.  Again, the zeroth order phase transition is a product of the fact that negative entropy black holes are regarded as unphysical, and hence discontinuities are introduced into the Gibbs free energy. 

As $\alpha$ increases further, we observe thermodynamic behaviour that is qualitatively similar to what we previously described. 
However, as $\alpha$ becomes larger, the zeroth order and first order phase transitions begin at larger values of the pressure.  The result is that at $\alpha \sim 12$ all phase transitions occur for pressures larger than the maximal pressure constraint.  

In the case of nonzero $\tq$ we observe no  {qualitatively different behaviour}.  The charge simply acts to shift the phase transitions in both $\alpha$ and $p$.   

\subsection{Thermodynamics in $D \ge 7$}

Employing the positivity of entropy condition, we find that for $k= \pm 1$ there are no critical points in {$D=7,8,9,10$} that have $p_c < p_+$ regardless of the value of $\tq$.  Furthermore, we find there to be no interesting phase phenomena present.  Accounting for the positivity of entropy, the Gibbs free energy displays two branches that meet at a cusp for all values of $\alpha$ and $\tq$.    

Based on the form of the equation of state we expect this situation to hold for all $D \geq 7$. Furthermore, 
 since the thermodynamics in quasitopological gravity and 3$^{rd}$-order Lovelock gravity are identical for $D \ge 7$, we expect the same results will hold if this analysis was performed for  3$^{rd}$-order Lovelock gravity with negative coupling.

\section{Conclusion}

Our investigation of the thermodynamic behaviour of black holes ({coupled to a $U(1)$ gauge field})
in cubic quasitopological gravity has yielded a number of interesting results.  We found that their thermodynamic behaviour 
was quite similar in a number of respects to their counterparts in  cubic Lovelock gravity, but also found a number of striking differences. 
Throughout  our study we have ensured that the parameters 
respect the physical criteria of positive black hole entropy and  stable ghost-free vacua via the implementation of various thermodynamic constraints.

As far as the similarities are concerned, we found that the equation of state in  $D\geq 7$ is identical to the cubic Lovelock case, and so $\mu>0$ black holes will exhibit phase behaviour similar to that already 
studied for this case  \cite{Frassino2014}.  
This means black holes in cubic quasitopological gravity exhibit the same novel behaviour,
particularly isolated critical points with non-standard critical exponents \cite{Dolan:2014vba}.

However the equation of state for the quasitopological case extends to $D=5$, unlike the cubic Lovelock case. Our investigation of this sector indicated that for spherical black holes there is a single critical point (with standard critical exponents) provided 
$\alpha \gtrsim 1.9909$.  In the hyperbolic case no critical phenomena were found.  

The preceding considerations hold for $\mu > 0$.   
For $\mu < 0$ our study of the thermodynamics of $D=5$ black holes proved to be particularly fruitful.  For $k=+1$ spherical black holes we  found examples of small/large/small black hole reentrant phase transitions in the uncharged case, and the presence of a triple point and a small/intermediate/large black hole phase transition reminiscent of the solid/liquid/gas phase transition in the charged case.  Although both of these features have been previously found in gravitational systems (first studied in the context of the Kerr-AdS solution in six dimensions \cite{Altamirano2013}), this work represents the first instance of these features observed in five dimensions. 
All of these were obtained for ghost-free vacua satisfying constraints of standard AdS asymptotics and positive entropy.
While we found an example of an isolated critical point, it occurred within a ghosty AdS vacuum.

For $k=-1$ hyperbolic black holes we found examples of small/large/small black hole reentrant phase transitions as well as small/large/small/large black hole multiple reentrant phase transitions.  This latter case represents the first time a multiple reentrant phase transition has been observed in a five dimensional gravitational system, having previously only been found in seven and higher dimensions \cite{Frassino2014}. 
We found all instances of isolated critical points to occur for negative pressures.

We found no new or interesting thermodynamic phenomena in {$D=7,8,9$ or $10$} for negative quasitopological coupling. Based on the form of the equation of state, we expect this situation to hold for all $D\geq 7$ and for 3rd-order Lovelock theories with negative coupling.
 
Further thermodynamic territory remains to be mapped out in quasitopological gravity. Our  analysis could be extended to negative $\alpha$ for $\mu > 0$. While this region of parameter space  contains black hole solutions and appears to have some critical behaviour, the 
analysis  is complicated by the fact that  one is not guaranteed to have AdS asymptotics for all branches.
Hence it could be that the thermodynamics corresponds to  a branch of a compact spacetime. 
To keep track of the asymptotic structure would require a nontrivial analysis of the
asymptotics for all branches as a function of the parameters $\tq$, $m$, and $\alpha$.

\acknowledgements 

We thank David Kubiz\v n\'ak for helpful advice and Maple resources.  This work was partially funded by NSERC through their Discovery, Vanier, and PGS programmes.

\clearpage
\providecommand{\href}[2]{#2}\begingroup\raggedright\endgroup

\end{document}